\documentclass[amsmath,amssymb,aps,prl,reprint,groupedaddress,superscriptaddress]{revtex4-2}

\usepackage{graphicx}
\usepackage{dcolumn}
\usepackage{bm}

\usepackage{floatrow}

\usepackage{hyperref}
\usepackage[mathlines]{lineno}
\usepackage{multirow}

\begin{document}


\preprint{}

\title{Observation of Traveling Breathers and Their Scattering in a
  Two-Fluid System}



\author{Yifeng Mao}
\email{yifeng.mao@colorado.edu}
\affiliation{Department of Applied Mathematics, University of Colorado, Boulder, CO 80309, USA}

\author{Sathyanarayanan Chandramouli}
\affiliation{Department of Mathematics, Florida State University, Tallahassee, FL 32306, USA.}

\author{Wenqian Xu}
\affiliation{Department of Applied Mathematics, University of Colorado, Boulder, CO 80309, USA}

\author{Mark Hoefer}
\affiliation{Department of Applied Mathematics, University of Colorado, Boulder, CO 80309, USA}


\date{\today}

\begin{abstract}

  The observation of traveling breathers (TBs) with large-amplitude
  oscillatory tails realizes an almost 50-year-old theoretical
  prediction \cite{kuznetsov_stability_1975} and generalizes the
  notion of a breather.  Two strongly nonlinear TB families are
  created in a core-annular flow by interacting a soliton and a
  nonlinear periodic (cnoidal) carrier. Bright and dark TBs are
  observed to move faster or slower, respectively, than the carrier
  while imparting a phase shift. Agreement with model equations is
  achieved. Scattering of the TBs are observed to be physically
  elastic. The observed TBs generalize to many continuum and discrete
  systems.

\end{abstract}


\maketitle



Fifty years ago, special, localized two-soliton bound state solutions
of the sine-Gordon equation were called ``breathers'' because they
incorporate two time scales: one associated with propagation and the
other associated with internal oscillation \cite{ablowitz1973method}.
The term breather has since been generalized to solutions of other
completely integrable systems such as the modified Korteweg-de Vries
(mKdV) and the focusing nonlinear Schr\"odinger (NLS) equations
\cite{ablowitz1981solitons}. However, the strict notion of localized
breather solutions appears to be limited to integrable
equations. Numerical simulations \cite{kudryavtsev1975solitonlike,
  ablowitz1979solitary, campbell1983resonance} and mathematical
analysis \cite{segur1987nonexistence, soffer1999resonances}
demonstrate that breathers in some non-integrable equations
necessarily display small oscillatory tails. Consequently, they have
been referred to as quasi-breathers \cite{ablowitz1981solitons} or
nonlocal solitary waves \cite{boyd1990numerical}, although, in some
instances, these oscillatory tails may be vanishingly small
\cite{kalisch_breather_2022}.

Small amplitude breathers can be approximated by bright soliton
solutions of the focusing NLS equation.  Consequently, breathers have
been observed in many situations such as water waves
\cite{pethiyagoda_spectrograms_2017,chabchoub_directional_2019},
internal waves \cite{grimshaw_modelling_2017}, nonlinear optical
\cite{mollenauer_experimental_1980,mandelik_observation_2003,boyd_nonlinear_2013}
and matter \cite{di_carli_excitation_2019,luo_creation_2020} waves,
rogue waves in these systems
\cite{kharif_rogue_2009,kibler_peregrine_2010,chabchoub2011rogue,narhi_real-time_2016,tikan_prediction_2022},
magnetic materials \cite{camley_nonlinear_2011}, and discrete systems
\cite{flach_discrete_2008,boechler2010discrete}.

Breathers with non-vanishing oscillatory tails have been studied
extensively in lattice systems where they are commonly referred to as
traveling intrinsic localized modes or \textit{traveling breathers}
\cite{iooss2005localized,flach_discrete_2008,james2018travelling} with
experimental observations of their existence in
\cite{sato2007driven,kim2015highly}. 
In the continuum setting, special solutions, originally called
solitary dislocations, were derived for the KdV equation as the
nonlinear superposition of a soliton and a cnoidal traveling periodic
wave in the seminal work of Kuznetsov and Mikhailov
\cite{kuznetsov_stability_1975}. This was achieved using inverse
scattering theory (IST) by introducing a discrete eigenvalue,
representing the soliton, in either the finite or semi-infinite gap of
the cnoidal wave spectrum for the Schr\"odinger operator associated
with KdV.  This solution is a traveling breather with large amplitude
oscillatory tails that
can be of either elevation (bright) or depression (dark) type
depending on the eigenvalue location in the semi-infinite gap or finite gap,
respectively. 
The cnoidal wave can also be considered a soliton lattice and so this
traveling breather can be viewed as a soliton-soliton lattice
interaction, generalizing the localized two-soliton bound state
breathers in, e.g., sine-Gordon and mKdV \cite{ablowitz1981solitons}.
The scattering theory for KdV traveling breathers indicates that they
interact with each other elastically, experiencing only a spatial
shift \cite{kuznetsov_stability_1975}.  Despite much ensuing analysis
\cite{belokolos_algebro-geometric_1994,gesztesy_m-kdv_1995,hu2012explicit,lou2015consistent,takahashi_integrable_2016,kuznetsov2017fermi,nakayashiki_one_2020,bertola2022partial,hoefer2022KDV,krichever1975potentials}
and corresponding physical interest, the observation and scattering of
strongly nonlinear traveling breathers in continuum mechanics is
lacking.

In this Letter, we create and scatter strongly nonlinear elevation
wave or bright (BB) and depression wave or dark (DB) traveling
breathers at the interface between two viscous fluids by the
overtaking interaction of a carrier wave with a soliton or
vice-versa. The traveling breather oscillatory tails---the carrier
background---are large amplitude.  The results experimentally prove
that this class of breathers can be interpreted as a nonlinear
superposition between a soliton and a cnoidal-like wave, realizing the
spectral interpretation of a traveling breather from IST developed
almost 50 years ago
\cite{kuznetsov_stability_1975,krichever1975potentials}.

Additionally, we measure traveling breather properties during free
propagation and scattering, finding that they retain their solitonic
character, i.e., interactions are physically elastic.  Unimodal and
bimodal geometries of internal oscillation are observed.  Qualitative
or quantitative features of individual traveling breathers and their
scattering agree with predictions from KdV traveling breather theory
or numerical simulations of a viscous two-fluid nonlinear model,
respectively.  Increasing traveling breather speeds imply decreasing
(BB) or increasing (DB) carrier phase shifts.  Slower (faster)
traveling breathers exhibit a negative (positive) spatial shift
post-interaction.  The observation of traveling breathers in this
model continuum system and their characterization are relevant to the
physics of nonlinear dispersive media more broadly.

The experiments are conducted in a tall acrylic column of
$5\text{cm}\times5\text{cm}\times180\text{cm}$, described in
\cite{mao2023experimental}.  The column consists of a pressure-driven
viscous core fluid with a free interface to a miscible, heavier
viscous reservoir fluid with a small core-to-reservoir fluid viscosity
ratio.  We precisely control the injection of the buoyant, interior
fluid in the low Reynolds number regime, which is convectively
unstable \cite{martin2009convective, selvam2009convective}, so that
straight conduits are established with constant injection.
Time-varying injection results in a conduit that exhibits an
azimuthally symmetric interface with cross-sectional area $A(z,t)$,
where $z > 0$ is the vertical spatial coordinate in the camera view
that is slightly above the injection site.  We identify $t=0$ as the
time at the initiation of interfacial imaging.  At the injection site,
the cross-sectional area satisfies the Hagen-Poiseuille flow law
$Q \propto A^2$ so it can be precisely controlled by temporally
varying the injection rate $Q$
\cite{scott1986observations,olson1986solitary}.

Viscous core-annular interfacial waves are modeled by the strongly
nonlinear conduit equation expressing mass conservation
\cite{olson1986solitary,lowman2013dispersive}
\begin{equation}
  A_t + \left(A^2\right)_z - \left(A^2 \left(A^{-1}A_t \right)_z
  \right)_z=0, 
  \label{eq:conduit}
\end{equation}
given in non-dimensional form where $z = \tilde{z}/L$,
$t = \tilde{t}/T$, $A = \tilde{A}/A_0$ for dimensional quantities
$\tilde{z},\tilde{t},\tilde{A}$ and $L$, $T$, $A_0$ are the fitted
vertical length, time, and cross-sectional area, respectively,
obtained from separate measurements of linear dispersive waves on the
background injection rate $Q_0 \propto A_0^2$ \cite{supplement}.  The
conduit equation \eqref{eq:conduit} and variants of it are models of
magma flow \cite{barcilon1986nonlinear} and channelized water flow in
glaciers \cite{stubblefield2020solitary}. The conduit equation is
apparently not integrable and can be reduced to the KdV equation in
the long-wavelength and small-amplitude regime \cite{supplement,
  harris2006painleve}.  Cnoidal-like waves in the form
$A(z,t) = g(\theta)$, $\theta = kz - \omega t$,
$g(\theta + 2\pi) = g(\theta)$ and solitons are generated by
evaluating numerically computed traveling wave solutions of the
conduit equation \cite{maiden2016modulations} at the injection site,
which determines the injection rate time-series
$Q(t) \propto g(-\omega t)^2$. The characterization and reliable
generation of soliton and cnoidal-like waves have, individually, been
reported in \cite{olson1986solitary, helfrich1990solitary,
  mao2023experimental}.


\begin{figure*}  
 \centering
  \includegraphics[width=0.24\textwidth]{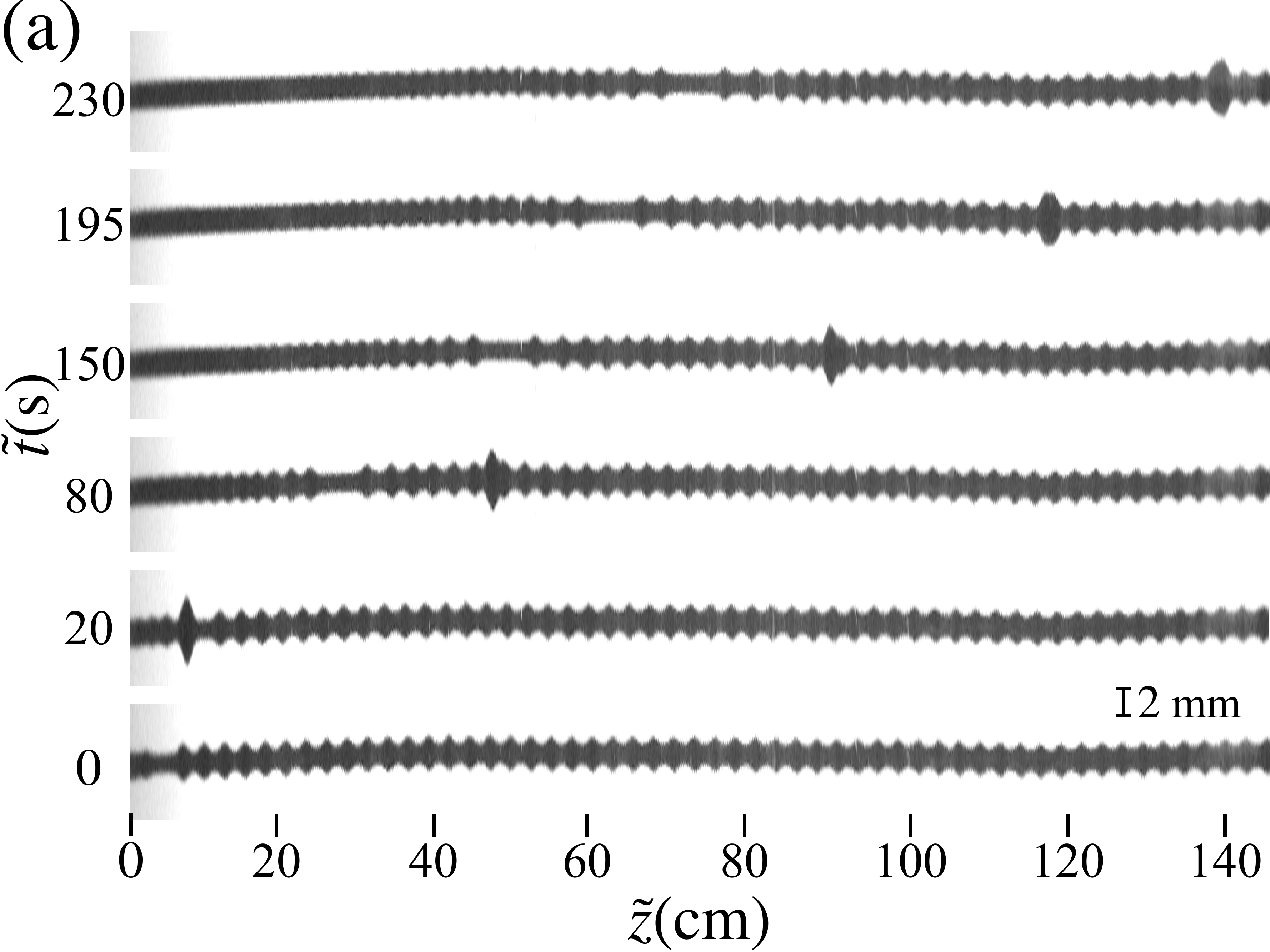}
  \includegraphics[width=0.24\textwidth]{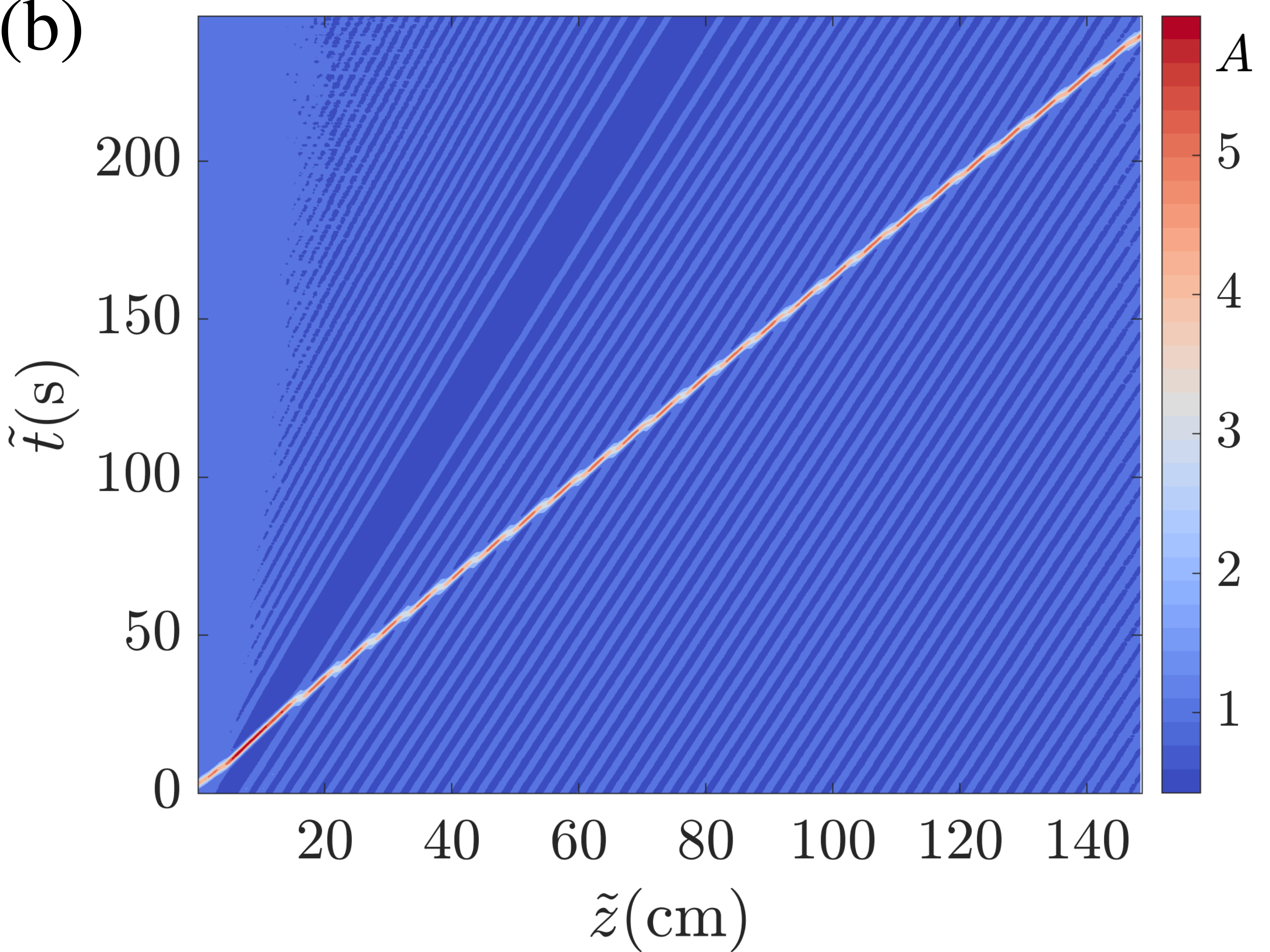} 
  \includegraphics[width=0.24\textwidth]{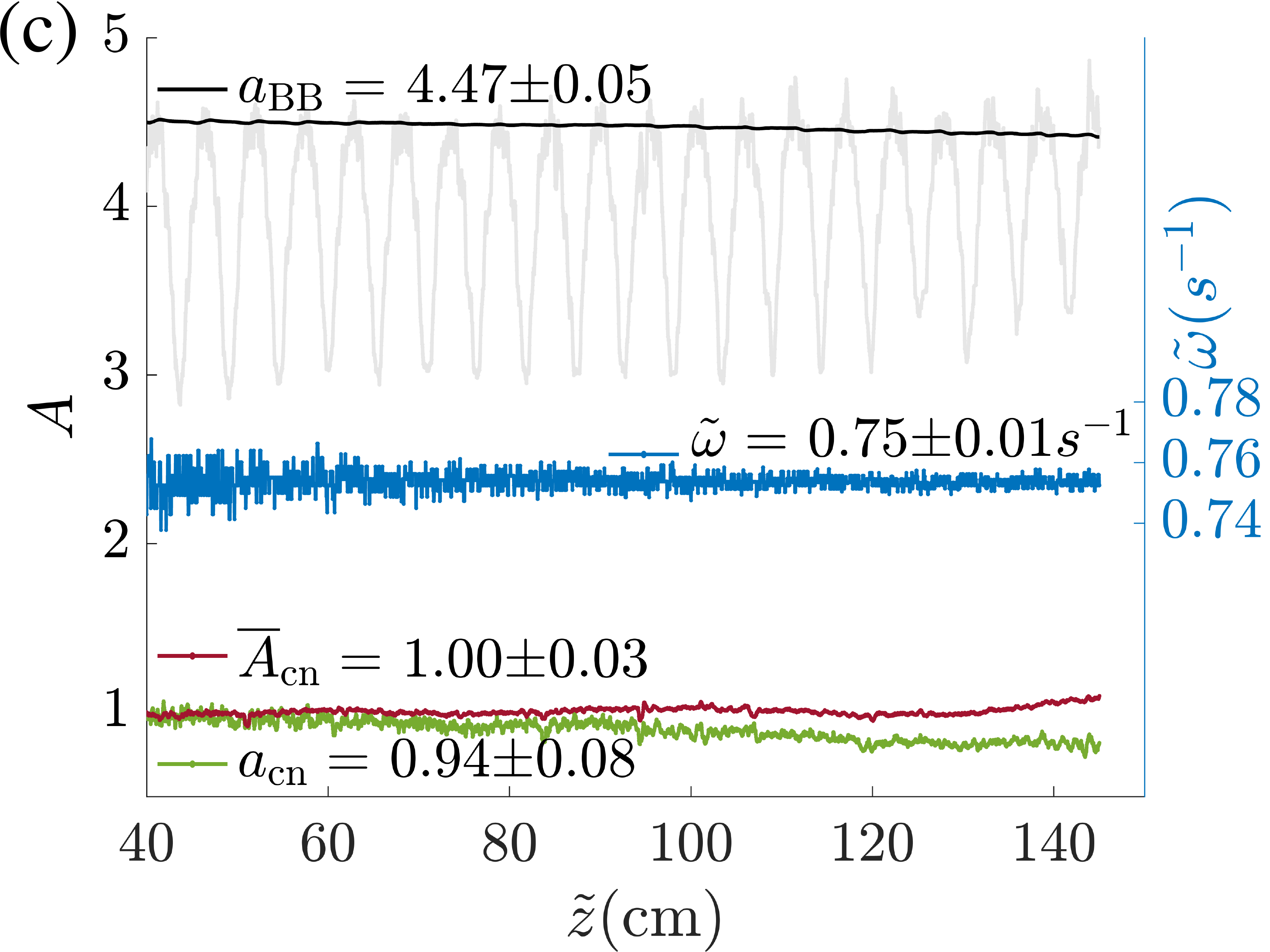}  
     \includegraphics[width=0.24\textwidth]{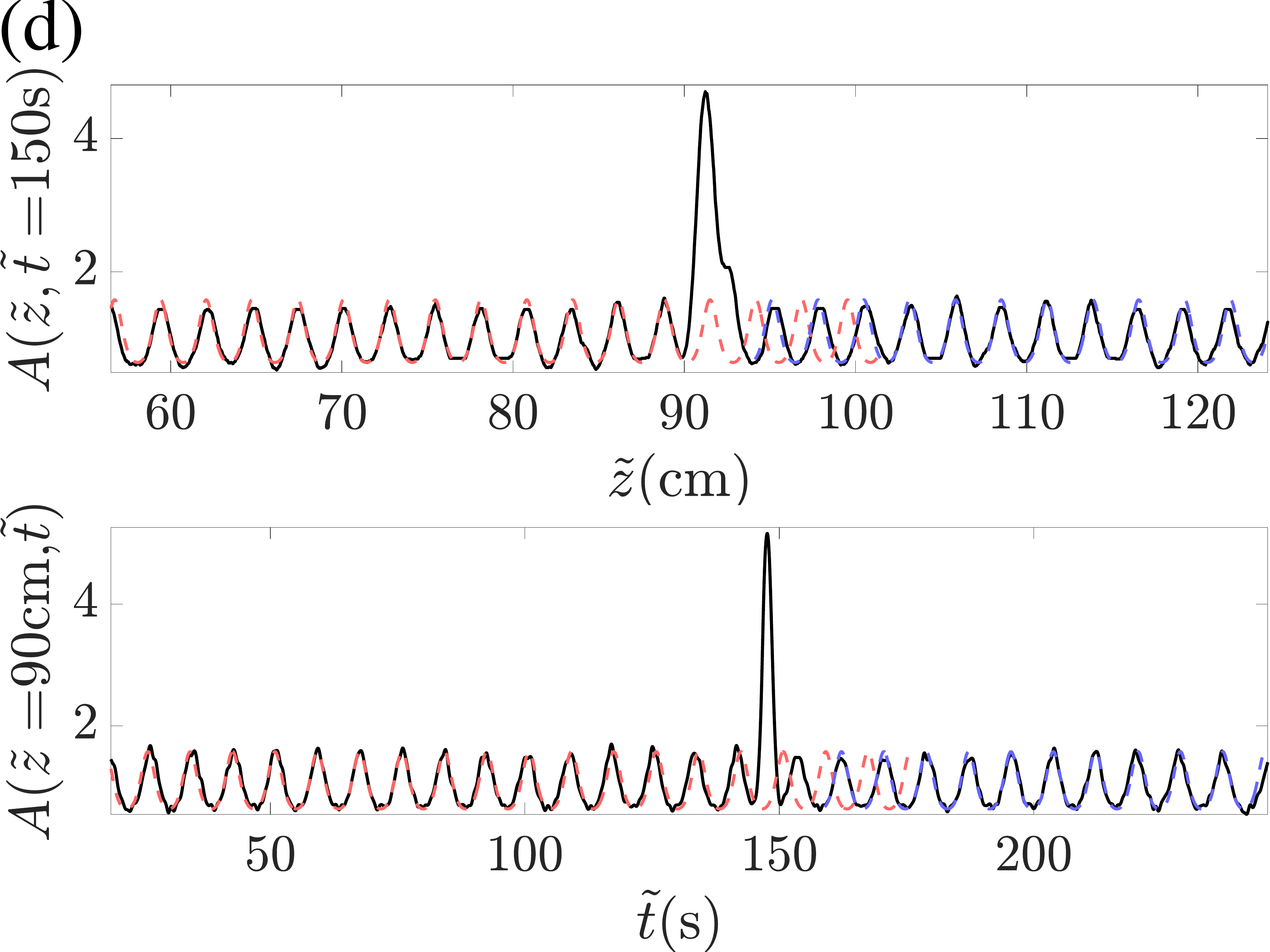}
    \caption{Bright traveling breather (BB) in a viscous fluid conduit formed from
    a soliton overtaking a cnoidal-like periodic traveling wave. (a)
    $90^{\circ}$ clockwise rotated time-lapse. (b) Space-time contour
    of the cross-sectional area. (c) Background cnoidal-like wave and
    BB measurements at different heights.  (d) Spatial and temporal
    profiles (black solid) and the cnoidal-like wave solution of the
    conduit equation \eqref{eq:conduit} corresponding to measured
    parameters (dashed).  }
  \label{fig:bright_timelapse}
\end{figure*}

Figure \ref{fig:bright_timelapse} displays the spatio-temporal
evolution of a soliton overtaking a cnoidal-like wave.  At $t = 0$, we
have prepared the conduit as a nonlinear cnoidal-like wave (the
\textit{carrier}) that abruptly terminates at the injection site to a
constant flow rate for $t > 0$. The carrier propagates upward for
positive $z$ and experiences a modulation region as it transitions
from periodic to a constant background
(Fig.~\ref{fig:bright_timelapse}a for $t > 20$s,
Fig.~\ref{fig:bright_timelapse}b) \cite{supplement}.  This modulated
region consists of a dispersive shock wave (DSW)
\cite{maiden2016observation}, a constant region whose value
$A_{\rm min}$ coincides with the minimum of the adjacent
carrier, 
and a carrier modulation
\cite{maiden2016observation,gavrilyuk2021singular}. The averaged
carrier's amplitude ($a_{\rm cn}$), mean ($\overline{A}_{\rm cn}$),
and dimensional angular frequency ($\tilde{\omega} = \omega/T$) are reported
in Fig.~\ref{fig:bright_timelapse}c.  The measured angular wavenumber
($\tilde{k} = k/L$) is $\tilde{k} = 2.35\pm0.01\text{cm}^{-1}$.  After
terminating the carrier, a large-amplitude soliton is injected,
which is transmitted through the DSW \cite{maiden2018solitonic} and
the constant region $A_{\rm min}$. Because the soliton speed on the
constant region $A_{\rm min}$ exceeds the carrier's phase speed
($\tilde{v}_{\rm ph} = \tilde{\omega}/\tilde{k}$), the soliton
overtakes the carrier and forms a new coherent structure.  The
amplitude of the coherent structure extracted from experiment
($a_{\rm BB}(z) = \max_t A(z,t) - A_{\rm min}$) oscillates with
propagation.  In Fig.~\ref{fig:bright_timelapse}c, we plot
$a_{\rm BB}(z)$, its envelope, and the carrier's properties
extracted from each spatial slice of Fig.~\ref{fig:bright_timelapse}b
to characterize propagation up the conduit.  The amplitude, mean, and
frequency of the carrier and the coherent structure's amplitude
envelope exhibit small fluctuations with height $z$.  We conclude that
the coherent structure is a BB consisting of the nonlinear
superposition of a soliton and a cnoidal-like wave.

Next, we extract the BB's maximum value for $\tilde{z}\gtrsim 40$cm in
Fig.~\ref{fig:bright_timelapse}b that follows a zig-zag path with
constant speed regularly interspersed between regions of rapid
acceleration/deceleration.  The BB speed across the entire zig-zag
path is $\tilde{v}_{\rm BB}=0.63\pm0.01$cm/s \cite{supplement}, which
exceeds the measured carrier phase speed
$\tilde{v}_{\rm ph} = 0.319\pm 0.005$cm/s.  This is consistent with
KdV BB solutions \cite{kuznetsov_stability_1975, hoefer2022KDV}.  A more thorough analysis of
the BB trajectory \cite{supplement} shows that it can also be viewed
as a soliton undergoing a sequence of phase shifts while interacting
with a soliton lattice that composes the cnoidal-like wave
\cite{kuznetsov_stability_1975, whitham_comments_1984}.

In Fig.~\ref{fig:bright_timelapse}d, we extract experimental time and
spatial slices of $A(z,t)$.  The measured BB's phase shift in space
and time $\Delta \theta \in (-\pi,\pi]$ is defined as the difference
between the left and right carrier phases.  We find that
$\Delta \theta_{\rm BBz} = 0.72\pi$ in space and
$\Delta \theta_{\rm BBt} = 0.75\pi$ in time
are comparable, as expected.  This BB with a positive phase shift is
consistent with large amplitude BB solutions of the KdV equation
\cite{hoefer2022KDV}.


\begin{figure*}
  \includegraphics[width=0.24\textwidth]{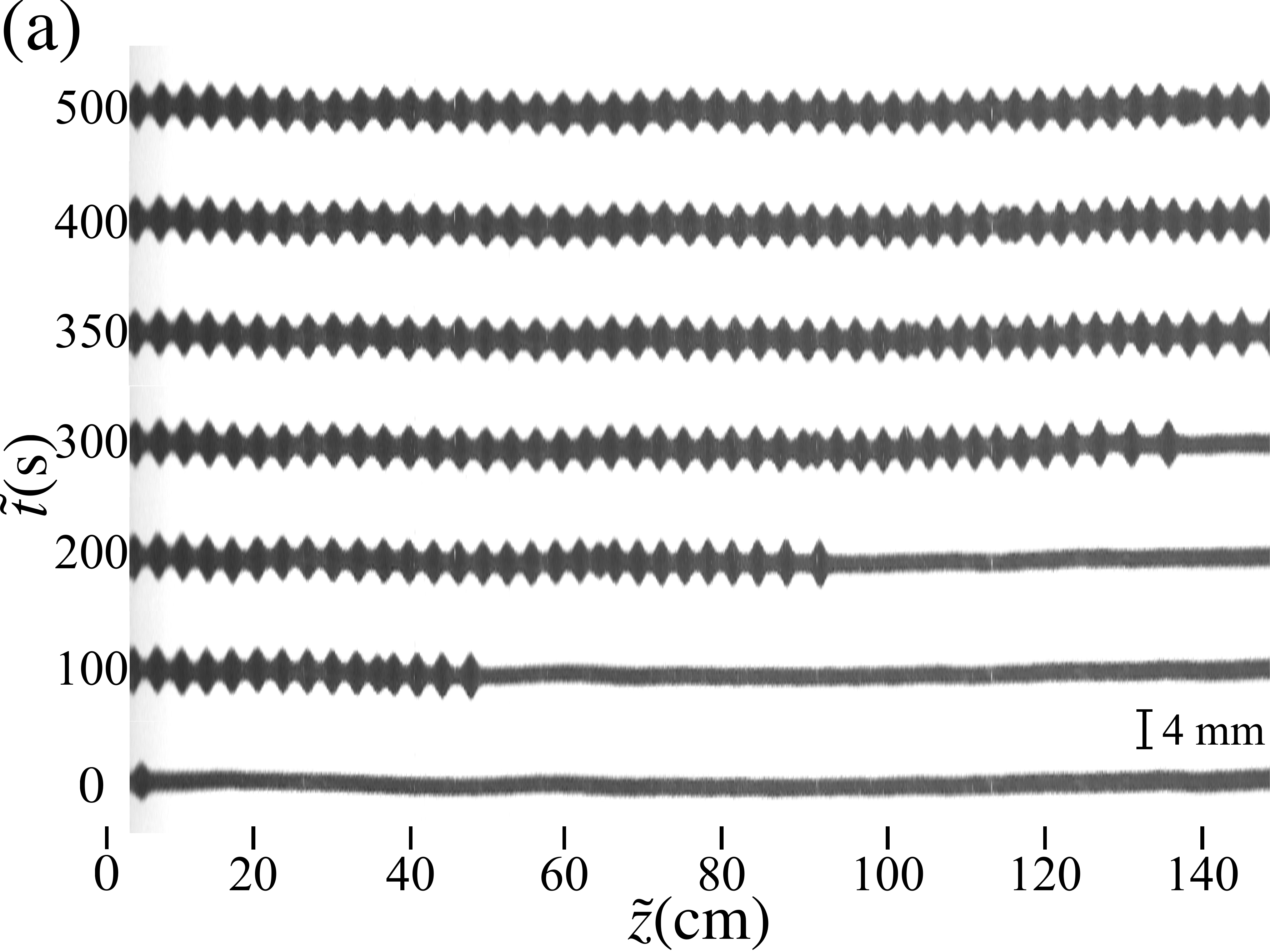}
  \includegraphics[width=0.24\textwidth]{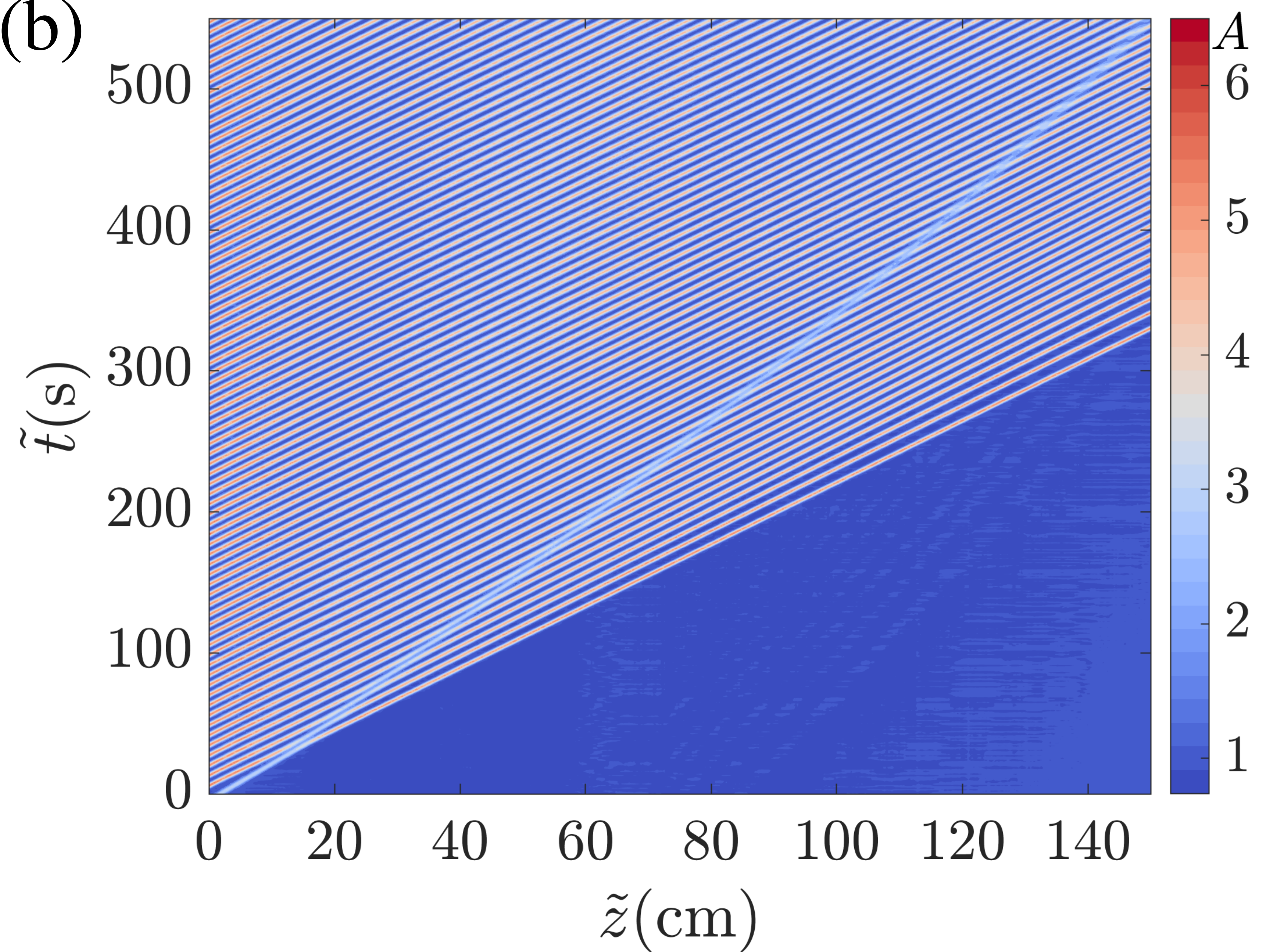} 
  \includegraphics[width=0.24\textwidth]{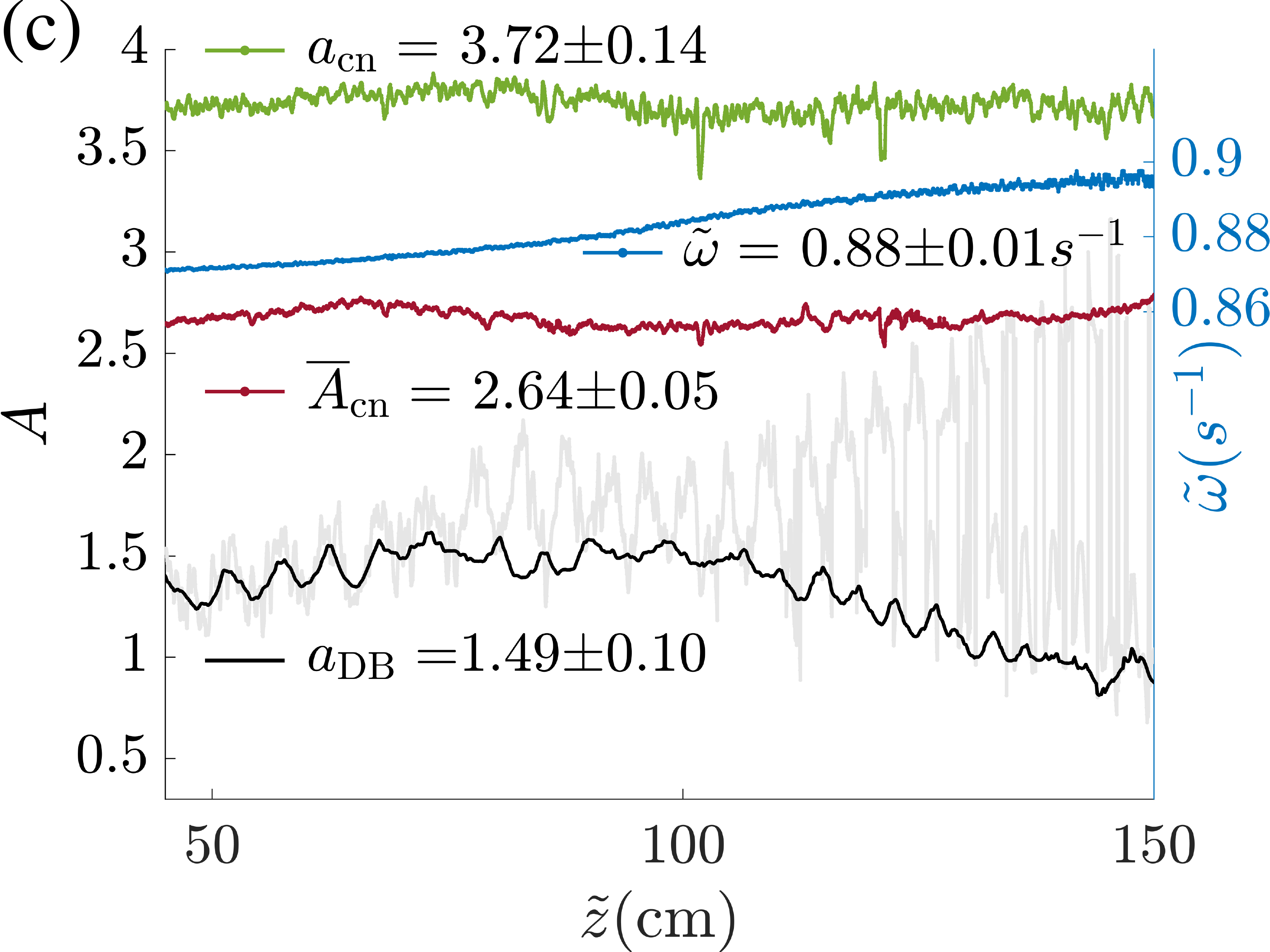}
  \includegraphics[width=0.24\textwidth]{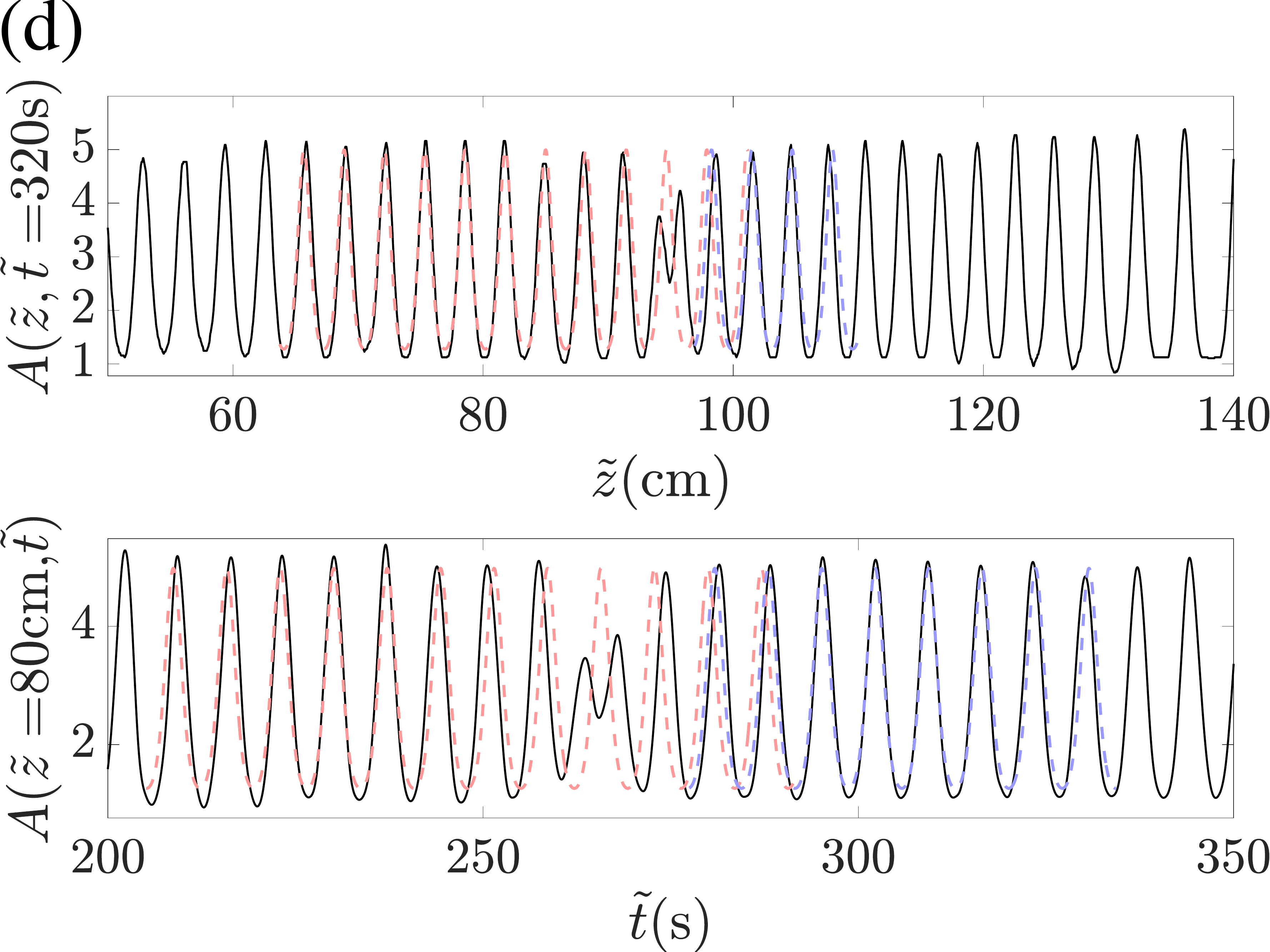}
  \caption{Absorption of a solitary wave by a larger-mean,
    larger-amplitude cnoidal-like periodic traveling wave resulting in
    a dark traveling breather. (a) $90^{\circ}$ clockwise rotated time-lapse
    images. (b) Space-time contour.  (c) Background cnoidal-like wave
    and DB measurements.  (d) Spatial and temporal profiles (black
    solid) and the cnoidal-like wave solution of \eqref{eq:conduit}
    corresponding to measured parameters (dashed). }
  \label{fig:dark_timelapse}
\end{figure*}

To generate a DB, 
the cnoidal-like carrier is set to overtake the soliton.  So that the
carrier fully absorbs the soliton, it is essential to increase the
mean flow rate of the carrier relative to the soliton.  This approach
is reported in Fig.~\ref{fig:dark_timelapse}a,b where, now,
soliton-cnoidal-like wave interaction results in a depression defect,
a DB.  For each fixed $z$, we define the DB amplitude $a_{\rm DB}(z)$
as the difference between the maximum of the carrier and the minimum
of the DB upper envelope.  The measured averaged carrier parameters
are given in Fig.~\ref{fig:dark_timelapse}c and
$\tilde{k} = 1.95\pm 0.03\text{cm}^{-1}$. In contrast to the BB case,
the carrier frequency shown in Fig.~\ref{fig:dark_timelapse}c
increases
up the conduit.  The 
carrier with a larger mean than the soliton leads to
more carrier fluctuations than in the BB case.  Near the top of the
conduit, we observe a 5\% increase in the carrier mean and a
corresponding change in other DB properties.  Generally, the injected
interior fluid is suspended at the top after rising through the
exterior fluid.  We attribute DB modulations for $\tilde{z} > 110$cm
to the slow diffusion of interior fluid from the top, which lowers the
exterior fluid's density $\rho_e$.  The observed 5\% mean increase can
be explained by a correspondingly small 2.5\% decrease in the exterior
to interior density difference via the Hagen-Poiseuille law in which
$\overline{A} \propto (\rho_e - \rho_i)^{-1/2}$
\cite{lowman2013dispersive}.  Note that a similar mean increase near
the top is observed for the BB in Fig.~\ref{fig:bright_timelapse}c.
Despite this, propagation of the DB is robust over a large portion of
the conduit $\tilde{z} \in [80,110]$cm where carrier and DB
fluctuations 
are modest.

The measured DB speed is $\tilde{v}_{\rm DB}=0.27\pm0.01$cm/s over
$\tilde{z} \in [80,110]\text{cm}$ while the carrier phase speed is
larger $\tilde{v}_{ph} = 0.45\pm 0.01$cm/s, a characteristic feature
of KdV DBs \cite{kuznetsov_stability_1975, hoefer2022KDV}. Figure \ref{fig:dark_timelapse}d
shows spatial, temporal slices and the determination of the carrier
phase shift. The phase shifts are
$\Delta \theta_{\rm DBz} = 0.24\pi$ in space and
$\Delta \theta_{\rm DBt} = 0.25\pi$ in time.


\begin{figure}
  \includegraphics[width=0.48\textwidth]{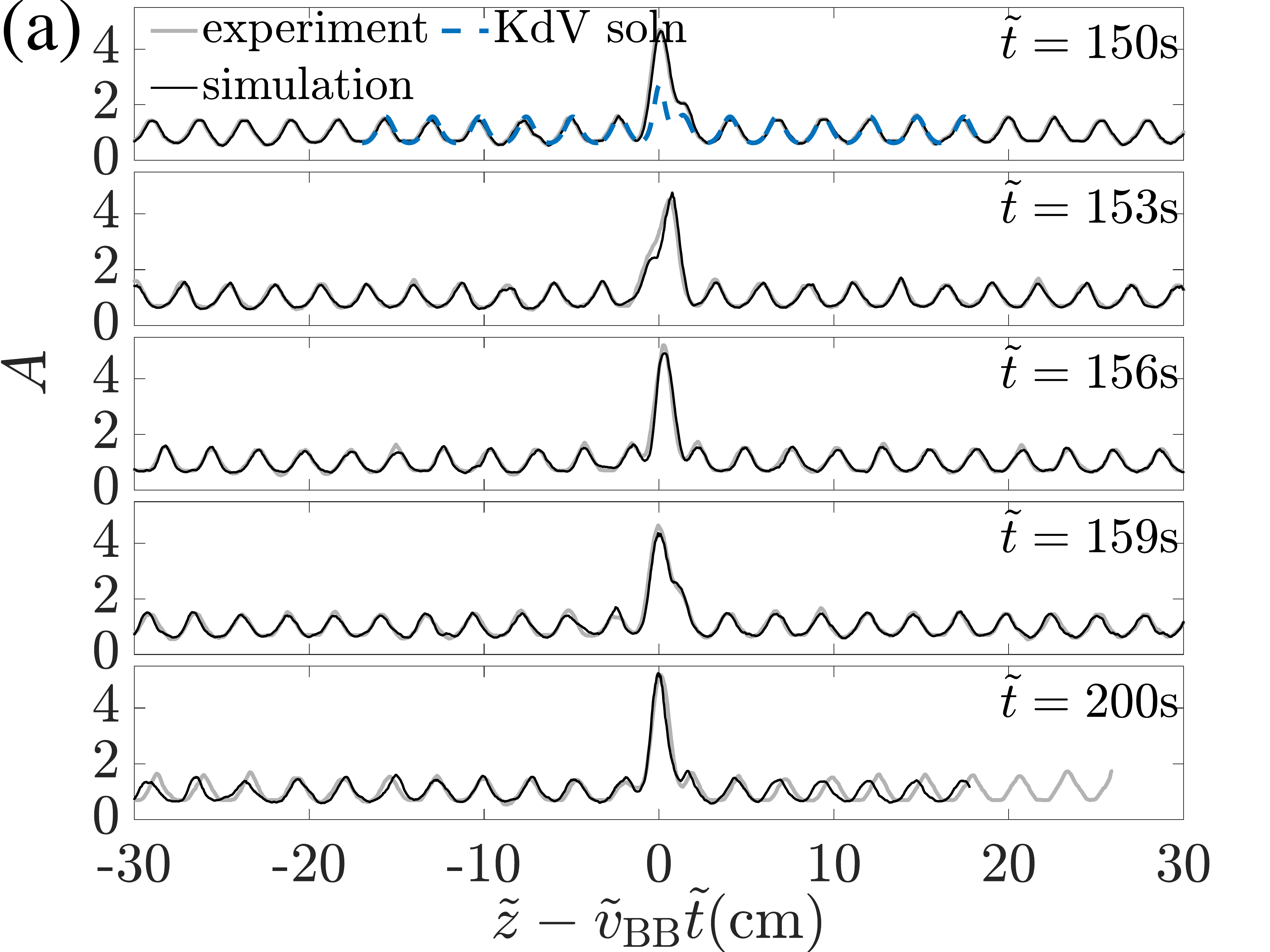}
  \includegraphics[width=0.48\textwidth]{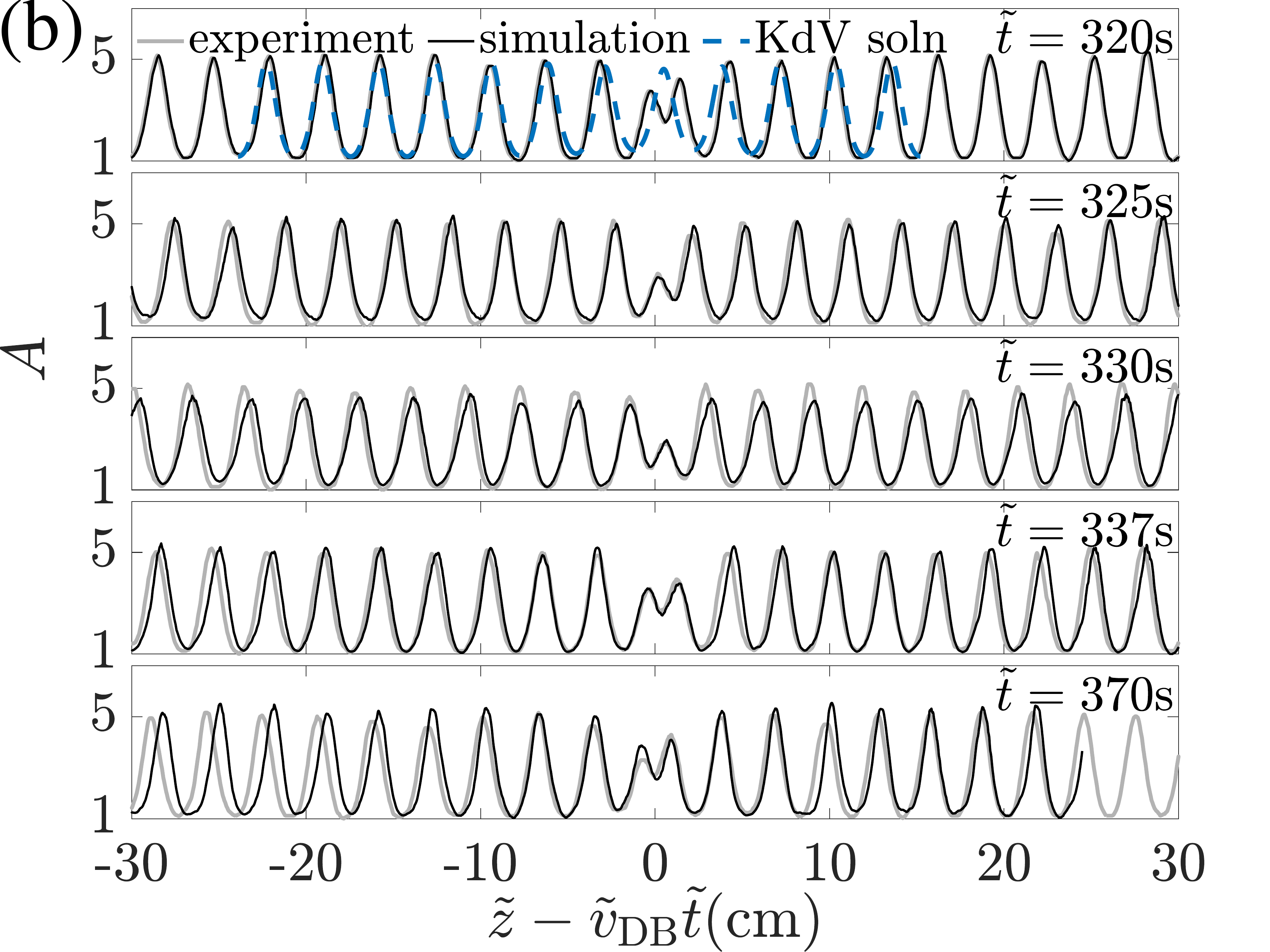}
  \caption{Experiment (light gray) compared with simulation of the
    conduit equation \eqref{eq:conduit} (black) with initial
    conditions from experiment, and the KdV traveling breather
    solution (blue).  (a) BB from Fig.~\ref{fig:bright_timelapse}d.
    (b) DB from Fig.~\ref{fig:dark_timelapse}d. The top four panels in
    (a,b) present the evolution over one period.}
\label{fig:sim}
\end{figure}


While qualitative features of observed BBs and DBs agree with KdV
theory, KdV is a quantitative model only for small amplitude
$a \ll 1$, long waves $k \ll 1$ subject to the dominant balance
$a \sim k^2$, where $a$ and $k$ are the nondimensional amplitude and
wavenumber of the carrier, respectively.  Since $a \gtrsim 1$ and
$k \approx 0.4$ in our experiments, the dominant balance and smallness
conditions are not satisfied.  Consequently, we perform numerical
simulations of the strongly nonlinear conduit equation
\eqref{eq:conduit} with periodic boundary conditions and an initial
condition extracted from the time sliced experimental sections
in Figs.~\ref{fig:bright_timelapse}d and
\ref{fig:dark_timelapse}d. Figure \ref{fig:sim} depicts a comparison
of the simulations, experiment, and a KdV traveling breather with
fitted carrier and phase shift in the comoving reference frame
\cite{supplement}.  In Fig.~\ref{fig:sim}a, an oscillation period of
the BB
($\tilde{T}_{\rm BB} = 2\pi/(\tilde{v}_{\rm BB} \tilde{k} -
\tilde{\omega}) \approx 9$s) is shown in the top four panels with the,
much smaller than observation, KdV BB solution.  Viewed as a
soliton-soliton lattice interaction, the BB exhibits the unimodal
interaction geometry according to the Lax categories of two-soliton
interactions
\cite{lax_integrals_1968,craig_solitary_2006,lowman2014interactions}.
For longer times, the simulation and experiment agree very well,
demonstrating the stability, robustness, and accuracy of conduit BBs
relative to observation.  Small carrier discrepancies are due to
unphysical periodic boundary conditions and the acknowledged diffusion
issues.  The DB in Fig.~\ref{fig:sim}b has period of oscillation
$\tilde{T}_{\rm DB} = 2\pi/(\tilde{\omega} - \tilde{v}_{\rm DB}
\tilde{k}) \approx 17$s, depicted in the top four panels.  The
dimensional time scale $T$ from eq.~\eqref{eq:conduit} used here is
5\% smaller than the nominal, measured $T$ extracted from separate
linear wave measurements \cite{supplement}.  We attribute this small
discrepancy to interior fluid diffusion and limitations of the conduit
equation as a model \cite{mao2023experimental}.  The KdV DB solution
differs from observation, which exhibits a bimodal interaction
according to the Lax categories \cite{lowman2014interactions}.  The DB
simulation closely tracks experiment subject to some carrier
discrepancies.


\begin{figure}
  \includegraphics[height=0.39\textwidth]{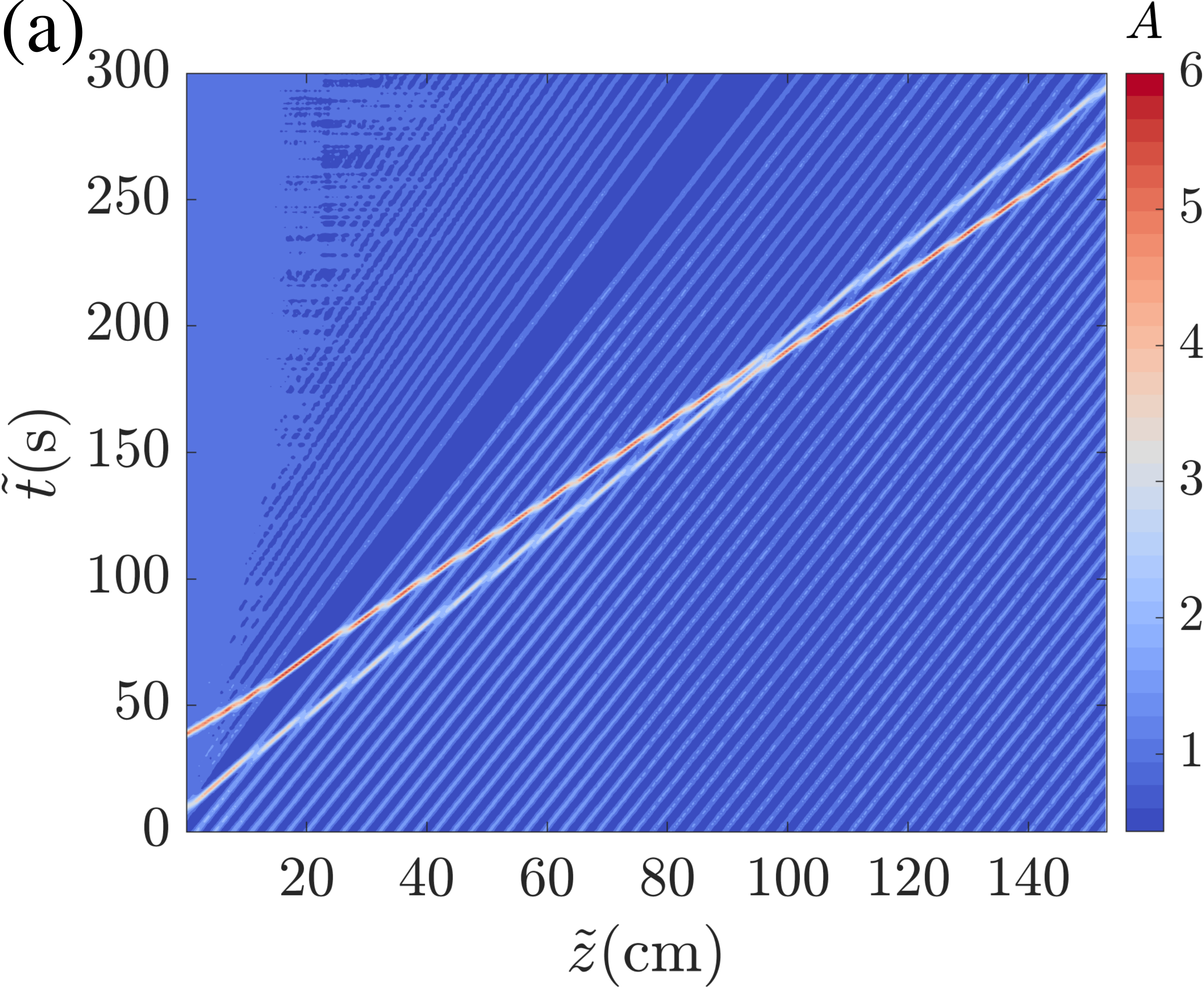}
  \includegraphics[height=0.39\textwidth]{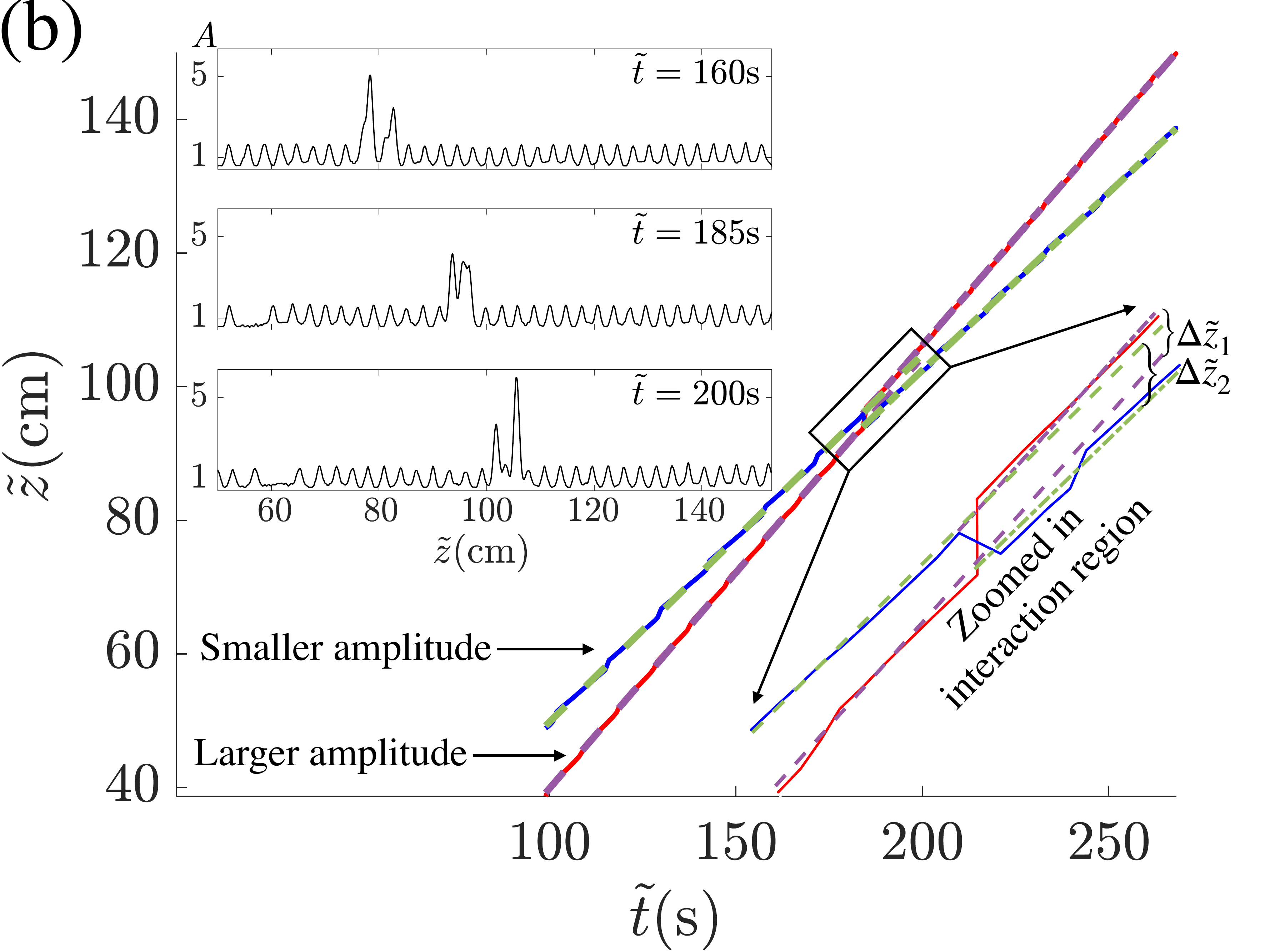}
  \\
  \includegraphics[height=0.39\textwidth]{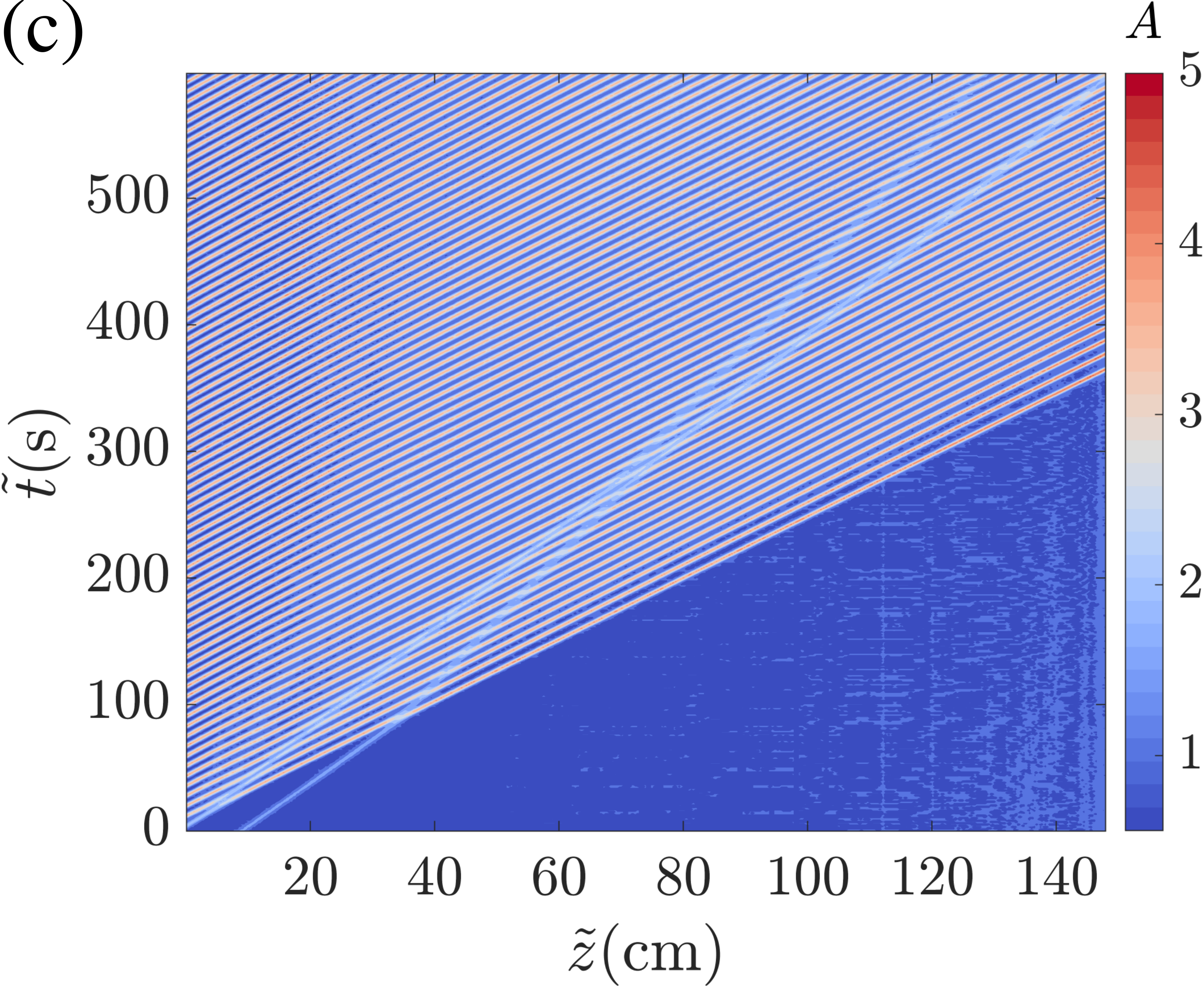}
  \includegraphics[height=0.39\textwidth]{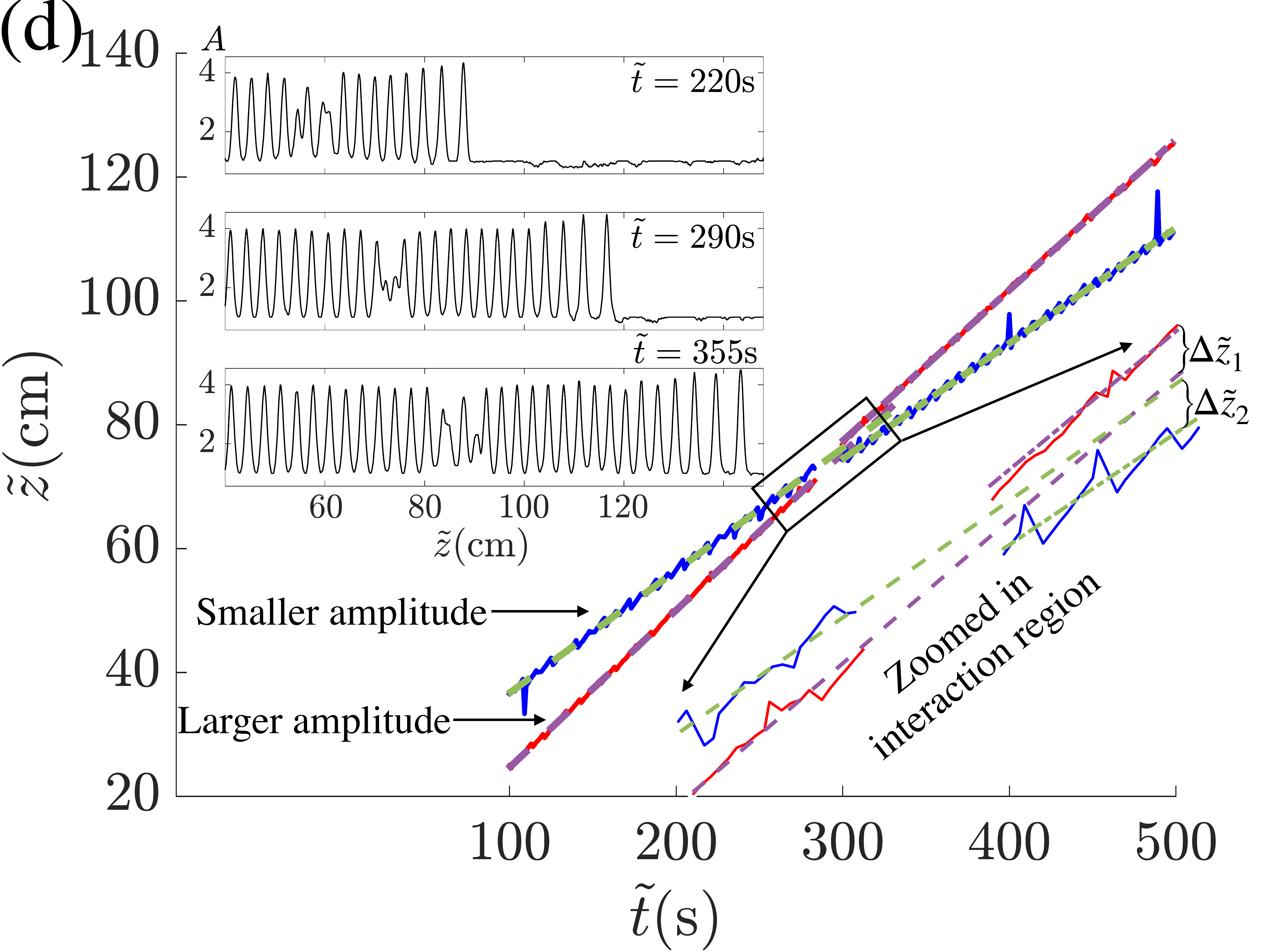} \\
  \includegraphics[height=0.39\textwidth]{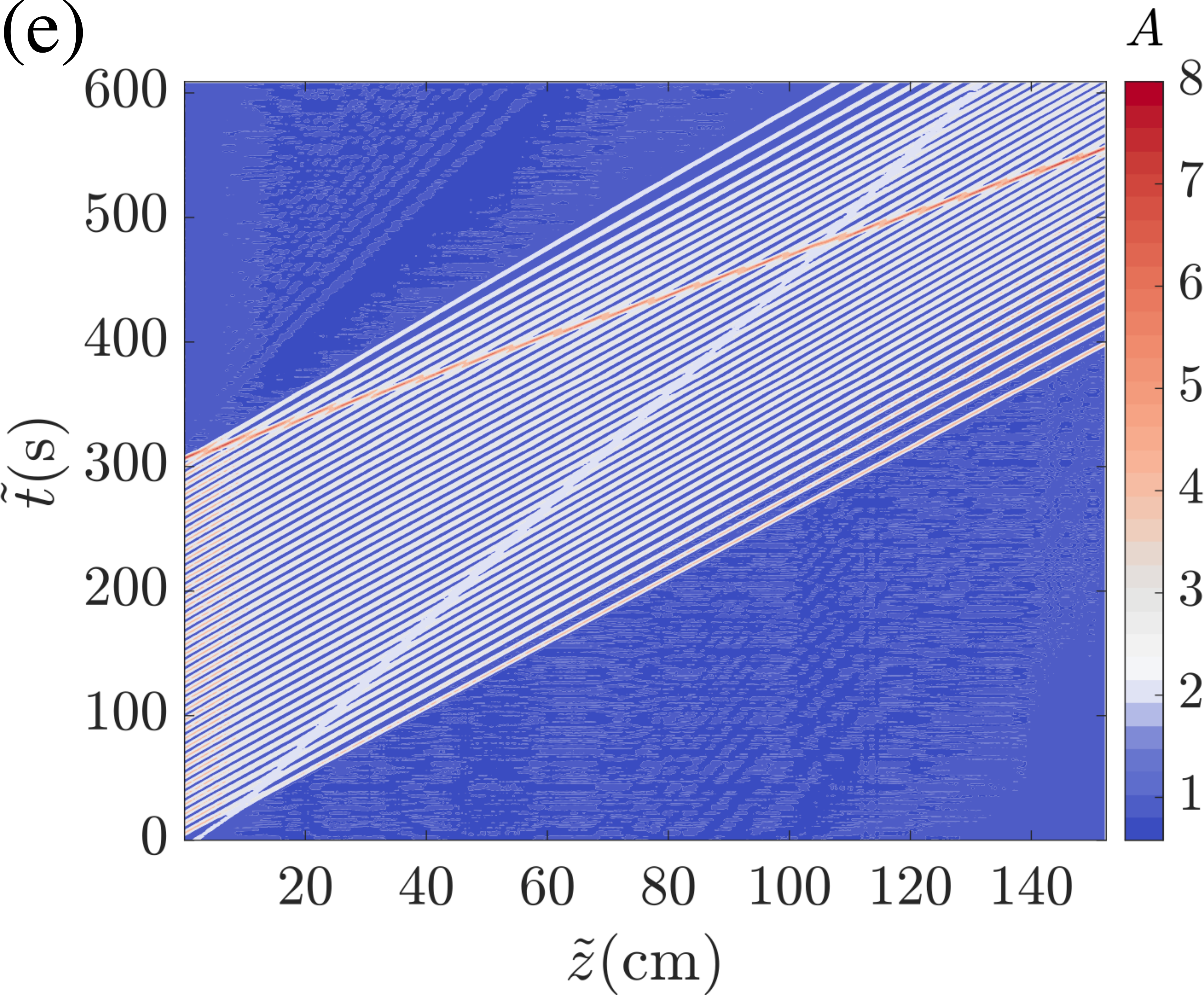}
  \includegraphics[height=0.39\textwidth]{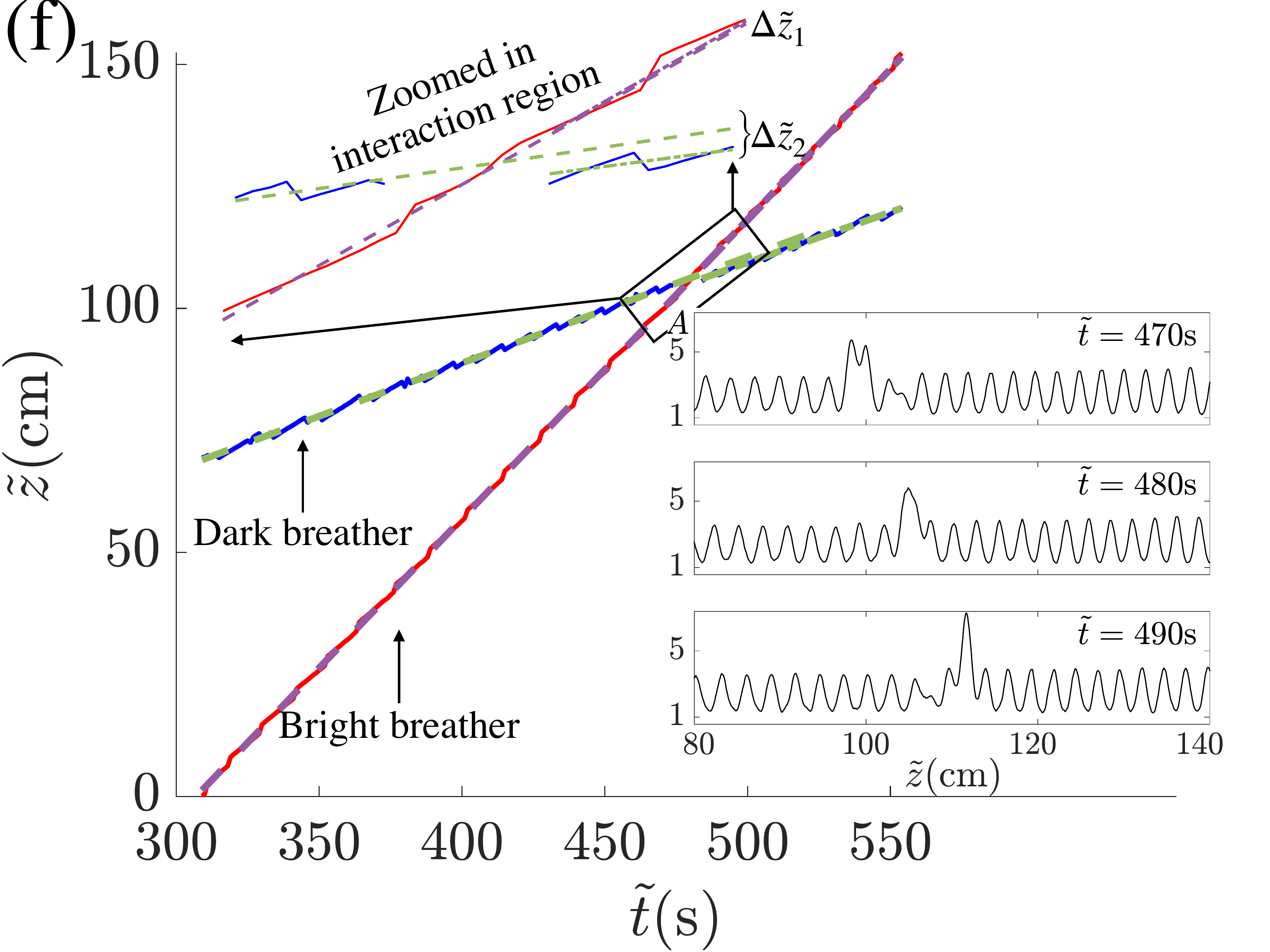}
  \caption{Observation of traveling breather scattering on a fixed
    carrier background. (a,c,e) Space-time contour revealing conduit
    dynamics. (b,d,f) Traveling breather trajectories (blue and red
    solid) in the $z$-$t$ plane with linear fits (dashed) and spatial
    shifts $\Delta \tilde{z}_j$. Subfigures provide representative
    spatial profiles at specific times.  (a,b) BB-BB scattering.
    (c,d) DB-DB scattering.  (e,f) BB-DB scattering.}
  \label{fig:breather_interactions}
\end{figure}

The reliable creation of single BBs and DBs in viscous fluid conduits
enables the investigation of traveling breather scattering.
We
create multiple traveling breathers on a fixed carrier background by
nonlinear superposing the periodic wavetrain with two solitons of
differing amplitude. Figure \ref{fig:breather_interactions}a depicts
the scattering of two BBs. The BB created from the larger-amplitude
soliton travels at a faster speed and overtakes the smaller amplitude
BB. Both BBs are observed to propagate coherently before and after
interaction, maintaining essentially the same
shape. Figure~\ref{fig:breather_interactions}b tracks BB peak
trajectories.  We separately fit the trajectories before and after
interaction with linear functions to derive BB speeds $\tilde{v}_j$
and spatial shifts $\Delta \tilde{z}_j$ ($j = 1,2$) occurring as a
result of the interaction (see Tab.~\ref{tab:interact_speed}).  For
each BB, the speeds pre and post interaction are the same, within very
small error tolerances, indicating that BB-BB scattering is physically
elastic. The faster (slower) BB experiences a positive (negative)
spatial shift.

Figure \ref{fig:breather_interactions}c presents DB scattering. The
larger amplitude DB originating from the larger amplitude soliton
moves faster than
the smaller amplitude, slower DB. Trajectories in
Fig.~\ref{fig:breather_interactions}d trace the positions of the DB
upper envelope minima. Again, DB spatial shifts are observed and DB
speeds pre and post interaction are conserved
(Tab.~\ref{tab:interact_speed}).  The faster (slower) DB undergoes a
positive (negative) spatial shift.

Since all BBs (DBs) are faster (slower) than the carrier, a BB must
overtake a DB for them to interact as shown in
Fig.~\ref{fig:breather_interactions}e.  The carrier must have finite
spatial extent in order to concurrently overtake and be overtaken by a
soliton to create a BB and DB.  The speed measurements of BB, DB
trajectories (Tab.~\ref{tab:interact_speed}) depicted in
Fig.~\ref{fig:breather_interactions}f again indicate that the
scattering is physically elastic.  
The DB spatial shift is negative but the positive BB spatial shift is
an order of magnitude smaller, hence the BB is scarcely affected by
the DB.  In all cases of observed traveling breather scattering, we
find that they are physically elastic and faster (slower) traveling
breathers experience a positive (negative) spatial shift due to the
interaction.  This observation is consistent with KdV traveling
breather scattering characterized in \cite{kuznetsov_stability_1975,
  bertola2022partial}.

In Fig.~\ref{fig:breather_interactions}, we launch traveling breather
pairs on the same carrier background.  By considering each traveling breather
independently, we can infer
their nonlinear dispersion relation.  In
Fig.~\ref{fig:breather_interactions}a, individual BB phase
shifts
after interaction are $\Delta \theta_{\rm BB1} = -1.08\pi$ and
$\Delta \theta_{\rm BB2} = -0.85\pi$.  Note that it is standard for a
phase shift below $-\pi$ to be shifted to $(0,\pi)$ by adding a $2\pi$
period.  However, by maintaining the phase shift sign, we can deduce
that an increasing BB speed corresponds to an increasing BB amplitude
and \textit{decreasing} BB phase shift. 
This is consistent with the KdV BB dispersion relation
\cite{hoefer2022KDV,supplement}.  In
Fig.~\ref{fig:breather_interactions}c, we measure the DB phase shifts 
post interaction to be $\Delta \theta_{\rm DB1} = 0.61\pi$ and
$\Delta \theta_{\rm DB2} = 0.52\pi$.  An increasing DB speed
corresponds to an increasing DB amplitude and phase shift.  For KdV
DBs, there are two branches of the DB nonlinear dispersion relation,
slow and fast associated with positive and negative phase shifts,
respectively \cite{hoefer2022KDV,supplement}.  Our experimental
observations are qualitatively consistent with the nonlinear
dispersion relation's slow branch of KdV DBs.

\begin{table*}[tb!]
\begin{tabular}{|c|c|c|c|c|c|c|}
\hline
\multirow{2}{*}{} & \multicolumn{3}{c|}{Traveling Breather 1} &
                                                                \multicolumn{3}{c|}{Traveling Breather 2} \\ \cline{2-7} 
& $\tilde{v}_1$ before (cm/s) & $\tilde{v}_1$ after (cm/s) & $\Delta \tilde{z}_1$ (cm) & $\tilde{v}_2$ before (cm/s) & $\tilde{v}_2$ after (cm/s) & $\Delta \tilde{z}_2$ (cm) \\ \hline
BB-BB & $0.65\pm0.01$ & $0.64\pm0.01$ & 1.54 & $0.55\pm0.01$ & $0.53\pm0.01$ & -2.69 \\ \hline 
DB-DB & $0.25\pm0.01$ & $0.24\pm0.01$ & 1.75 & $0.20\pm0.01$ & $0.19\pm0.01$  & -1.97 \\ \hline
BB-DB & $0.61\pm0.01$ & $0.62\pm0.01$ & 0.15 & $0.22\pm0.01$ & $0.21\pm 0.01$ & -1.22 \\ \hline
\end{tabular}
\caption{Speed and spatial shift measurements of traveling breather scattering.}
\label{tab:interact_speed}
\end{table*}


This work experimentally verifies that traveling breathers in a
continuum system can be generated from the interaction of solitons and
periodic cnoidal-like traveling waves.  While continuum traveling
breather theory has been developed for integrable systems, our
observations and numerical simulations of the two-fluid system suggest
that traveling breathers exist in non-integrable systems as well.
Traveling breathers appear to be natural, practical and common
phenomena that occur in continuum environments accompanied by a
possibly large amplitude oscillatory background, distinguishing them
from localized breathers.  While spectral data analysis techniques
have been developed to extract this type of information from complex
signals using IST-based nonlinear Fourier analysis
\cite{costa_soliton_2014,bruhl_comparative_2022}, we have physically
realized this nonlinear superposition principle with a simple
method. The scattering of traveling breathers is observed to be
physically elastic (energy conserving) while exhibiting a spatial
shift.  This work lays the foundation for exploring traveling
breathers in other continuum systems such as optics, fluids, condensed
matter and anywhere nonlinear waves occur.




The authors would like to thank the Isaac Newton Institute for
Mathematical Sciences, Cambridge, for support and hospitality during
the programme \textit{Dispersive hydrodynamics: mathematics,
  simulation and experiments, with applications in nonlinear waves},
where work on this paper was discussed.  This work was supported by
NSF DMS-1816934.


%

\end{document}



\preprint{}

\title{Supplemental Material for ``Observation and Scattering of Traveling Breathers in a
  Two-Fluid System"}


\author{Yifeng Mao}
\email{yifeng.mao@colorado.edu}
\affiliation{ Department of Applied Mathematics, University of Colorado, Boulder, CO 80309, USA}

\author{Sathyanarayanan Chandramouli}
\affiliation{Department of Mathematics, Florida State University, Tallahassee, FL 32306, USA.}

\author{Wenqian Xu}
\affiliation{ Department of Applied Mathematics, University of Colorado, Boulder, CO 80309, USA}

\author{Mark Hoefer}
\affiliation{ Department of Applied Mathematics, University of Colorado, Boulder, CO 80309, USA}


\date{\today}


\maketitle

\section{Experimental setup and method}

\begin{figure}[tb!]
\CenterFloatBoxes 
\begin{floatrow} 
\ffigbox[\FBwidth]
{\caption{Schematic of the experimental apparatus.}\label{fig: setup}}
{\includegraphics[width=0.4\textwidth]{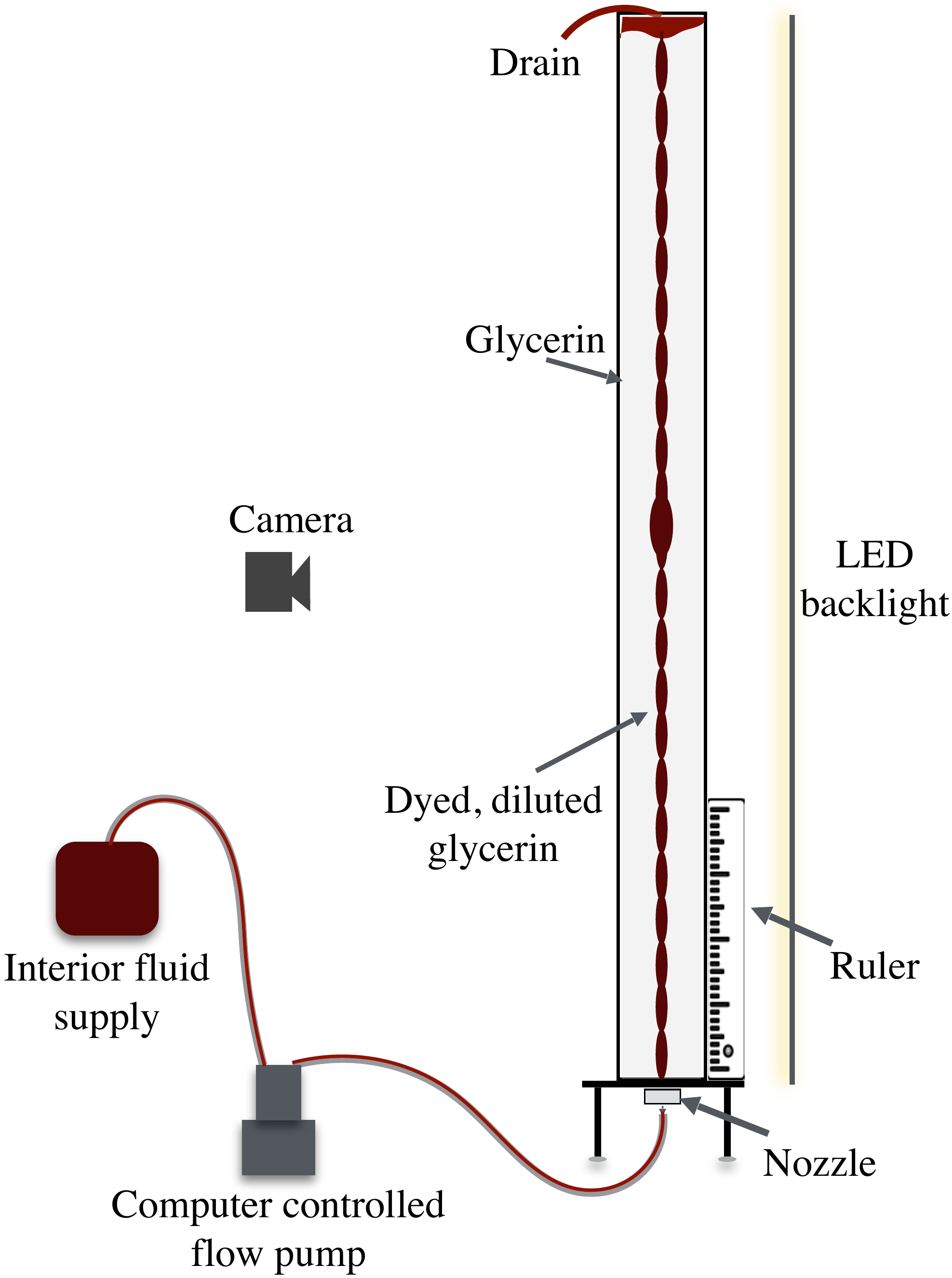}}

\killfloatstyle
\ttabbox[\Xhsize]
{
\rule{0.4\textwidth}{0.4pt}
\begin{tabular}{cc} 
$\mu^{(i)}$ &  $0.045$ Pa s\\
$\mu^{(e)}$ & $1.10$ Pa s  \\
$\epsilon$ & 0.041 \\
$\rho^{(i)}$ & 1.21 g cm$^{-3}$\\
$\rho^{(e)}$ & 1.26  g cm$^{-3}$\\
$Q_0$ & 1.00 cm$^3$ min$^{-1}$\\
$D_0$ & 0.56 cm\\
$D_{\rm out}$ & 5.00 cm\\
$Re^{(i)}$ & 0.21 \\
$Re^{(e)}$ & 0.0087\\
Camera resolution & 39.36 pix cm$^{-1}$\\
\end{tabular}
\rule{0.4\textwidth}{0.4pt}
}
{\caption{Example fluid properties (measured in the experiment of Fig.~1): viscosities $\mu^{(i,e)}$, viscosity ratio $\epsilon$, densities $\rho^{(i,e)}$, background flow rate $Q_0$, associated conduit diameter $D_0$ computed by Poiseuille's law, outer wall diameter $D_{\rm out}$, and Reynolds numbers $Re^{(i,e)}$ for interior and exterior fluids.}\label{tab: fluid_property}}
\end{floatrow}
\end{figure}

The experiments are conducted in a $180$ cm tall square acrylic column with a cross-sectional area of $5\times5$ cm$^2$ (see Fig. \ref{fig: setup} for a schematic illustration). The column is filled with high-viscosity pure glycerin as the exterior fluid, and the interior fluid, made of a mixture of glycerin ($\sim 70\%$), water ($\sim 20\%$), and black food coloring ($\sim 10\%$), which is less viscous and dense, is injected into the column using a computer-controlled piston pump. Table \ref{tab: fluid_property} reports nominal fluid properties. The interior fluid rises to the top and forms a floating thin layer that very slowly diffuses. We focus on the dynamics of the two-Stokes fluids' interface away from the top and consider the wavemaker problem with only one boundary condition at the injection site. Prior to imaging, a straight conduit is created by the steady injection rate
\begin{equation*}
	Q^{(i)}(t) = Q_0 q(t),  \quad q(t)=1,
\end{equation*}
where the superscript $(i)$ represents quantities for the interior fluid, $q(t)$ is a time-dependent dimensionless boundary condition, and $Q_0$ is a constant injection rate. We obtain various interfacial wave dynamics by programming the injection rate $q(t)$.

The conduit interfacial waves are imaged by a high-resolution digital camera with a sample rate of $2$ Hz for Fig.~1 of the main text and $1$ Hz for the remaining datasets. The camera is positioned slightly above the injection boundary and captures the interfacial waves in a fully established conduit. We identify the starting point in the camera view as $z=0$ and the imaging starting time as $t=0$. Conduit diameters $D$  are extracted by measuring the local extrema of centered differences of grayscale images in the direction normal to the interface. Conduit diameters approximately follow the Hagen-Poiseuille relation $D=\alpha Q^{(i)1/4}$ \cite{whitehead1975dynamics, scott1986observations}, where $\alpha= \left(2^7 \mu^{(i)} \right)^{1/4} / \left( \pi g \Delta \right)$, $\mu^{(i)}$ is the interior viscosity, $g$ is the acceleration due to gravity  and $\Delta=\rho^{(e)}-\rho^{(i)}$ is the difference in the exterior and interior densities. However, as reported in \cite{maiden2016observation}, the conduit diameter in the upper portion of the fluid column is found to be slightly larger than that near the bottom, which leads to changes in the interfacial wave profiles as discussed in the main text.

The normalized conduit cross-sectional area $A$ is obtained as $A=(D/D_0)^2$, where $D_0$ is the constant conduit diameter at the steady injection rate $q_0$. The initial-boundary value conditions for the wavemaker problem in terms of the conduit interfacial area $A$ are
\begin{align*}
	A(z,0) &= 1, \quad z\geq0, \\
	A(0,t) &= f(t), \quad t\geq0,
\end{align*}
where $f(t)=q(t)^{1/2}$.

Determination of the nondimensionalization scales $L,U$ and $T=L/U$ utilizes the method discussed in \cite{mao2023experimental}. After setting up each experiment, we generate trials of linear periodic waves in the fluid conduit and measure their dimensional wavenumber $\tilde{k}$ (rad/cm) and frequency $\tilde{\omega_0}$ (rad/s). Fitting $(\tilde{k},\tilde{\omega})$ with the conduit equation linear dispersion relation 
\begin{equation*}
	\tilde{\omega} = \frac{2U\tilde{k}}{1+L^2\tilde{k}^2},
\end{equation*} 
we obtain the vertical length scale $L$ (cm) and velocity scale $U$ (cm/s). The corresponding time scale is $T=L/U$ (s). The resulting scales for each experiment of the main text are reported in Table \ref{tab:carrier_parameters}.

\section{Generation of solitons and cnoidal-like periodic traveling waves}

Viscous fluid conduit interfacial waves have been shown to exhibit solitons that stably propagate along the conduit with constant speed and possess physical elasticity (speed conservation) after interaction with another soliton \cite{olson1986solitary,helfrich1990solitary}. Experimental generation of solitons in a conduit uses pre-computed soliton solutions of the conduit equation $f(t)=a_{\text{soli}} (-c_{\text{soli}}(t-t_0))$, where $c_{\text{soli}} =(1+a_0^2(2\ln a_0-1))/(a_0-1)^2$ is the wave speed, $a_0=a_{\rm soli}(0)$ is the soliton amplitude, and $a_{\rm soli}(z)$ is the soliton profile.

Conduit periodic traveling wave solutions satisfying $A(z,t)=g(\theta), \theta=kz-\omega t, g(\theta+2\pi)=g(\theta)$ for $\theta\in \mathbb{R}$ exhibit cnoidal-like patterns in the strongly nonlinear regime. In experiments, we use pre-computed conduit periodic traveling wave solutions $f(t)=g(-\omega t)$ to launch waves from the boundary. The wave propagating into a quiescent, straight conduit is subject to a developing dispersive modulation region. It is stabilized after sustained, periodic injection, at which time a periodic conduit is established. Investigation of its reliable creation and agreement with the conduit equation nonlinear dispersion relation have been realized in \cite{mao2023experimental}. Table \ref{tab:carrier_parameters} presents the averaged cnoidal-like carrier wave parameters extracted over a window of each reported dataset. Windows of measurement are chosen to focus on the periodic region solely. We scaled the bright traveling breathers (BBs) to have unit mean since the soliton and the carrier wave are on the same mean. On the other hand, we did not perform such scaling for the dark traveling breathers (DBs) since the DBs were intentionally created by interacting two waves with different means. The cnoidal-like wavetrain is subject
to a larger wave amplitude $a=A_{\max} - A_{\min}$ and a larger wave
mean $\bar{A}$.
The wavenumber is obtained as $k=2\pi/{l}$, where $l$ is the averaged wavelength in the spatial domain, and the frequency is $\omega=2\pi/p$, where $p$ is the averaged wave period in the temporal domain.

\begin{table*}[tb!]
\begin{tabular}{|c|c|c|c|c|c|c|c|c|}
\hline 
\multirow{2}{*}{Fig} & \multicolumn{6}{c|}{Carrier wave}  & \multicolumn{2}{c|}{Scales}   \\
\cline{2-9} 
  & region & $a$ & $\omega$ & $k$ & $\bar{A}$ & $k_n(a,\omega,\bar{A})$ & $L$ (cm) & $T$ (s)  \rule[0.8ex]{0pt}{2ex}  \\   \hline
  Fig. 1b & $z\in[40,100]$cm, $t\in[1,50]$s & $0.94\pm0.08$ &  $0.71\pm0.05$ &  $0.42\pm 0.02$ & $1.00\pm0.03$ & 0.43 & $0.18\pm0.01$ & $0.95\pm0.06$ \\ \hline
  Fig. 2b & $z\in[80,110]$cm, $t\in[400,500]$s & $3.72\pm0.14$ &  $1.21\pm0.08$ &  $0.31\pm 0.02$ & $2.64\pm0.05$ & 0.33 & $0.16\pm0.01$ & $1.38\pm0.09$\\ \hline
  Fig. 4a & $z\in[100,140]$cm, $t\in[1,100]$s & $0.96\pm0.07$ &  $0.68\pm0.04$ &  $0.39\pm 0.02$ & $1.00\pm0.04$ & 0.40 & $0.18\pm0.01$ & $0.95\pm0.06$ \\ \hline
  Fig. 4c & $z\in[40,70]$cm, $t\in[300,500]$s & $2.98\pm0.19$ &  $1.22\pm0.08$ &  $0.36\pm 0.02$ & $2.31\pm0.04$ & 0.39 & $0.19\pm0.01$ & $1.55\pm0.10$ \\ \hline
  Fig. 4e & $z\in[100,120]$cm, $t\in[400,450]$s & $2.52\pm0.16$ &  $1.25\pm0.07$ &  $0.37\pm 0.02$ & $2.40\pm0.05$ & 0.36 & $0.15\pm0.01$ & $1.43\pm0.08$ \\
  \hline
\end{tabular}
\caption{Nondimensional carrier wave parameters for each reported experiment in the main text. Column 2: selected windows for measurements. Columns 3-6: measured wave parameters. Column 7: numerically
computed solution $k_n$ at measured $(a,\omega,\bar{A})$, compared with measured $k$. Columns 8,9: nondimensionalization vertical length and time scales.  }
\label{tab:carrier_parameters}
\end{table*}

\section{Speed of the bright traveling breather}

\begin{figure}[tb!]
    \includegraphics[width=0.7\textwidth]{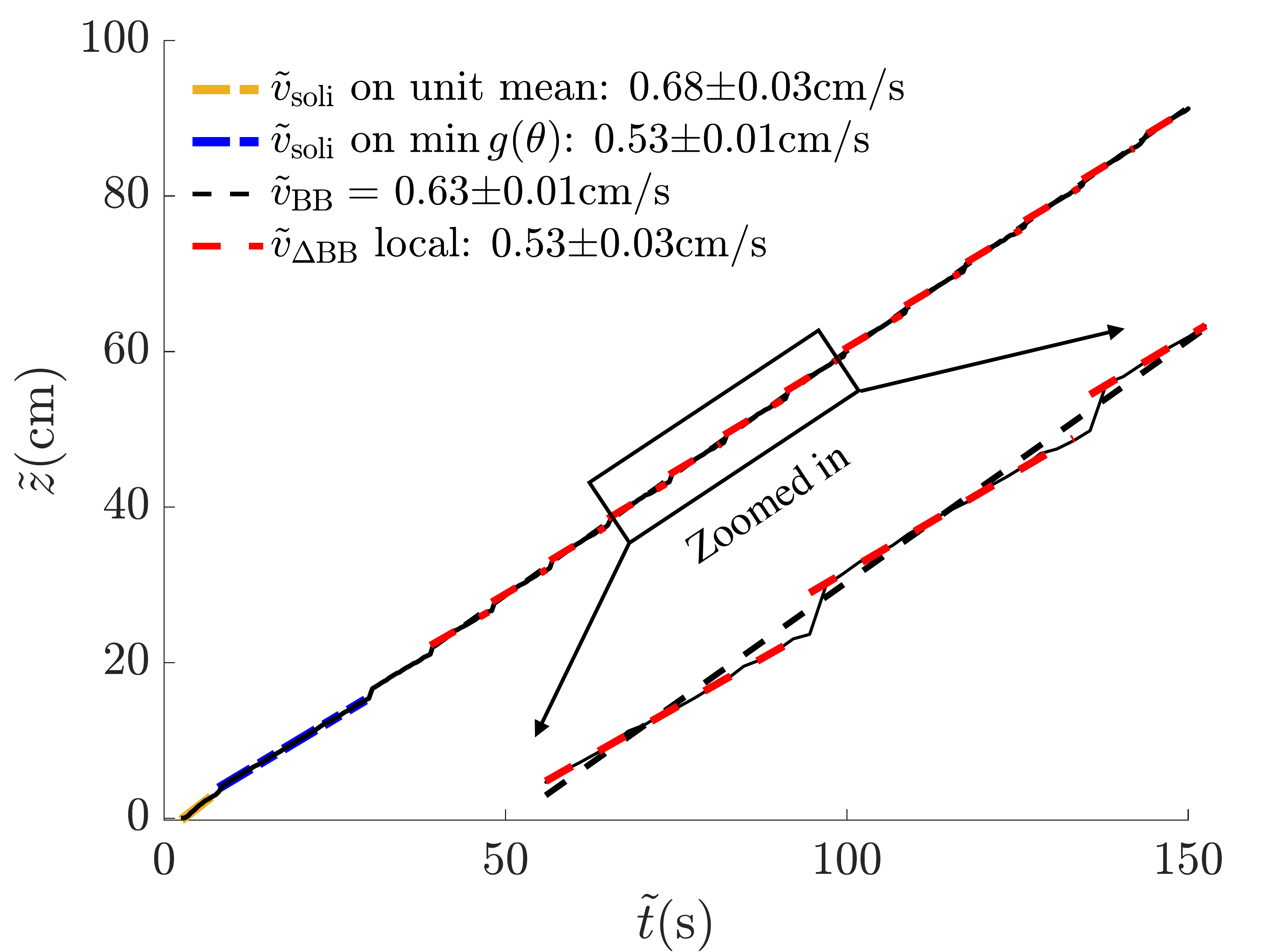}
\caption{Comparison of speeds for the bright traveling breather (BB) in Fig.~1b of the main text and the soliton before interacting with the carrier.  The space-time trajectory is fitted with linear functions for each sector, and the colors correspond to the different states of the soliton and the BB.}
\label{fig:bright_oscil_speed}
\end{figure}

The bright traveling breather (BB) in Fig.1b of the main text consists of a soliton overtaking a cnoidal-like periodic carrier wave, both of which are on unit mean. We now investigate the BB speed, compared with the associated soliton speed before interacting with the cnoidal-like carrier. In Fig.~\ref{fig:bright_oscil_speed}, we track the trajectory of the BB by extracting the position of the breather peak at each time slice. 

We consider the trajectory in three regions. The fitted green curve in Fig.~\ref{fig:bright_oscil_speed} presents the speed of the soliton on unit mean. The soliton then propagates into a constant state $A_{\rm min}$ that connects the unit mean and the minimum of the adjacent, localized carrier wave $A(z,t)=g(\theta)$, $\theta=kz-\omega t$. The speed of soliton propagation in this region is fitted in blue in Fig.~\ref{fig:bright_oscil_speed}. A BB is constructed after $\tilde{t}\approx 40$s when the soliton penetrates the carrier, tracked in black. As explained in the main text, the BB follows a zig-zag  path with constant speeds in between the spatial shifts. We, therefore, obtain the breather's \textit{local speed} $\tilde{v}_{\Delta {\rm BB}}$ by fitting the local trajectories in between each spatial shift. The speed of the BB across the entire zig-zag path is $\tilde{v}_{\rm BB}=0.63\pm0.01$cm/s, faster than the averaged local speed $\tilde{v}_{\Delta {\rm BB}}=0.53\pm0.03$cm/s because of the forward spatial shifts. We also observe that the soliton on unit mean (green) has a faster speed than that on the constant, lower state $A_{\rm min}$ (blue), as expected. More importantly, the soliton on the constant state $A_{\rm min}$ has speed $\tilde{v}_{\rm soli}=0.53\pm 0.01$cm/s, the same as the averaged BB local speed $\tilde{v}_{\Delta {\rm BB}}$, implying that the periodic carrier wave can be interpreted as a soliton train and the interaction of the soliton with the carrier consists of free propagation followed by an elastic interaction spatial shift periodically repeated.

\section{Comparison of conduit traveling breathers with KdV breather solutions}
\label{sec:breath-nonl-disp}

In \cite{kuznetsov_stability_1975, hoefer2022KDV}, two varieties of exact, traveling breather solutions of the Korteweg-de Vries (KdV) equation - bright and dark - are obtained, representing soliton-cnoidal wave interactions. These breathers impart a phase shift to the cnoidal wave. Herein, we compare both bright and dark traveling breathers from conduit experiments with the KdV traveling breather solutions by selecting a KdV solution that possesses the same cnoidal background and the same phase shift as the experimental data. Before that, we first need to derive the KdV equation from the conduit equation in the weakly nonlinear, long-wavelength regime.

The conduit equation
\begin{equation} 
    A_t + (A^2)_z - (A^2(A^{-1}A_t)_z)_z = 0, \label{eq:conduit}
\end{equation}
can be reduced to the KdV equations for
$T = \delta^{1/2}t$, $Z= \delta^{1/2} z$,
and
$A(z,t) = 1+\delta u(Z,T)$, $\delta\ll1$. 
Insertion of this ansatz into (\ref{eq:conduit}) and keeping the
leading and first-order terms results in the Benjamin–Bona–Mahony (BBM) equation
$u_T + 2u_Z + 2 \delta uu_Z - \delta u_{ZZT}= 0$.
Since $u_T=-2u_Z+\mathcal{O}(\delta)$,
$u_{ZZT}=-2u_{ZZZ}+\mathcal{O}(\delta)$ so that
$u_T + 2u_Z + 2\delta u u_Z + 2\delta u_{ZZZ} = 0$
to the same order of accuracy.  Hence, both the BBM and KdV
equations are asymptotically equivalent to the conduit equation
\eqref{eq:conduit} in the weakly nonlinear, long wavelength
regime.  Taking $\delta=1$, we arrive at the unscaled KdV equation 
\begin{equation} 
    u_t + 2u_z + 2uu_z + 2u_{zzz}=0. \label{eq:KdV_2}
\end{equation}
Letting $u(z,t) = 6v(x,\tau)$, where $x=z-2t$ and $\tau=2t$, \eqref{eq:KdV_2} is scaled into the normalized form 
\begin{equation}
v_\tau + 6vv_x+v_{xxx}=0 \label{eq:KdV_norm}
\end{equation}
as given in \cite{hoefer2022KDV}, which possesses the normalized cnoidal solution $v(x,\tau) = 2m \text{cn}^2(x-c_0 \tau,m)$, where $m$ is the elliptic parameter of the Jacobi elliptic cosine function and $c_0=4(2m-1)$ is the speed. Know that the KdV equation is subject to the scaling symmetry $ \Tilde{v} = \mu^2 v$, $\Tilde{x}=\frac{x}{\mu}$, $\Tilde{\tau} = \frac{\tau}{\mu^3}$, where $\mu>0$ and a Galilean boost so that the solution can be shifted to zero mean. We obtain the scaled, shifted cnoidal wave solution of the KdV equation \eqref{eq:KdV_2} on zero mean 
\begin{align} 
    u(z,t) = -6 \overline{u} + 12 m \mu^2 \text{cn}^2 \left[ \mu (z- 2( 1-6\overline{u} + c_0\mu^2 )t ), m \right],
    \label{eq:conduit_soln}
\end{align} 
where $\overline{u}=\frac{2m\mu^2}{K(m)}\int_{0}^{K(m)}\text{cn}^2(x,m)dx = 2\mu^2 \left(E(m)/K(m)-(1-m)\right)$. In addition, the conduit solution $A(z,t)=1+u(z,t)$ admits the scaling $A^*= \bar{A} A$, $z^*= \bar{A}^{1/2}z$, $t^* =\bar{A}^{-1/2}t$
that leaves \eqref{eq:conduit} invariant, where $\bar{A}>0$. Here we use asterisks to refer to the solution on mean $\bar{A}$.  Applying
this scaling invariance to \eqref{eq:conduit_soln}, we obtain the approximate cnoidal solution of the conduit equation
\begin{equation} 
    A^*(z^*,t^*) = \bar{A}(1-6\overline{u})  + a \text{cn}^2 \left[ \frac{\sqrt{a}}{\bar{A} \sqrt{12m}} \left( z^*- 2\left( \bar{A} - 6 \bar{A} \overline{u} + \frac{a(2m-1)}{3  m}\right) t^* \right), m \right], \label{eq:conduit_cnoidal2}
\end{equation} 
where $\overline{u}=\frac{a}{6\bar{A}m}\left(E(m)/K(m)-(1-m)\right)$. The solution \eqref{eq:conduit_cnoidal2} has mean $\bar{A}$, amplitude $a$, wavenumber $k=\frac{\pi\sqrt{a}}{K(m) \bar{A} \sqrt{12m}}$ and phase speed $C_0=2\left( \bar{A} - 6 \bar{A} \overline{u} + \frac{a(2m-1)}{3  m}\right)$. The corresponding frequency is given by $\omega=C_0 k$. We now fit \eqref{eq:conduit_cnoidal2} to experiment by measuring $(a,\bar{A},k)$ to determine $m$.

The top panels in Fig.~3 of the main text present comparisons of the
BB and DB from experiment and the KdV traveling breather solutions
approximated in the same way. In what follows, we provide the
theoretical details from weakly nonlinear KdV theory as predictions
for the experimental observations.  The normalized KdV equation
\eqref{eq:KdV_norm} admits BB and DB solutions in the form
\cite{hoefer2022KDV}
\begin{equation*}
    v(x,\tau) = 2\left[ m-1 + \frac{E(m)}{K(m)} \right] + 2 \partial_x^2 \log \nu(x,\tau),
\end{equation*}
where the $\nu-$function for the BB is given by 
\begin{equation*}
    \nu(x,\tau) := \Theta(x-c_0\tau + \alpha_b) e^{\kappa_b(x-c_b\tau+x_0)} + \Theta(x-c_0\tau - \alpha_b) e^{-\kappa_b(x-c_b\tau+x_0)}
\end{equation*}
with breather speed $c_b>c_0$, inverse width $\kappa_b$ and phase shift $-2\alpha_b\in(-2K(m),0]$, and the $\nu-$function for the DB is given by 
\begin{equation*}
    \nu(x,\tau) := \Theta(x-c_0\tau + \alpha_d) e^{-\kappa_d(x-c_d\tau+x_0)} + \Theta(x-c_0\tau - \alpha_d) e^{\kappa_d(x-c_d\tau+x_0)},
\end{equation*}
with breather speed $c_d<c_0$, inverse width $\kappa_d$ and phase shift $2\alpha_d\in[0,2K(m))$. $x_0$ is a spatial shift of the traveling breather relative to the carrier.
In Fig.~3a of the main text for the BB comparison,
to match the measured nondimensional wave parameters
$(a,\Bar{A},k)=(0.94,1.00,0.42)$ requires $m=0.82$ in the cnoidal
solution \eqref{eq:conduit_cnoidal2}. The BBs are on the same carrier
with phase shift $\Delta\theta_{\rm BB}=-1.28\pi$. The amplitude of
the KdV BB is nearly half of the conduit BB.  The measured carrier
phase speed is $v_{ph}=1.68$, compared to the predicted phase speed of
the cnoidal solution $C_0=1.74$. The first time displayed in Fig.~3b of the main text has carrier wave parameters $(a,\Bar{A},k)=(3.72,2.64, 0.31)$ and phase shift $\Delta\theta_{\rm DB}=0.24\pi$ for both the experimental DB and the KdV DB solutions. Fitting $(a,\bar{A},k)$ to \eqref{eq:conduit_cnoidal2} determines $m=0.83$. Similar to the BB comparison, the KdV DB has a smaller amplitude than the conduit DB. The experimental DB exhibits a bimodal structure during a period of oscillation, but the KdV DB is unimodal \cite{lowman2014interactions}. The measured and predicted cnoidal phase speeds are $v_{ph}=3.94$ and $C_0=4.33$, respectively.

\begin{figure}
    \centering
    \subfloat[]{\includegraphics[height=0.35\textwidth]{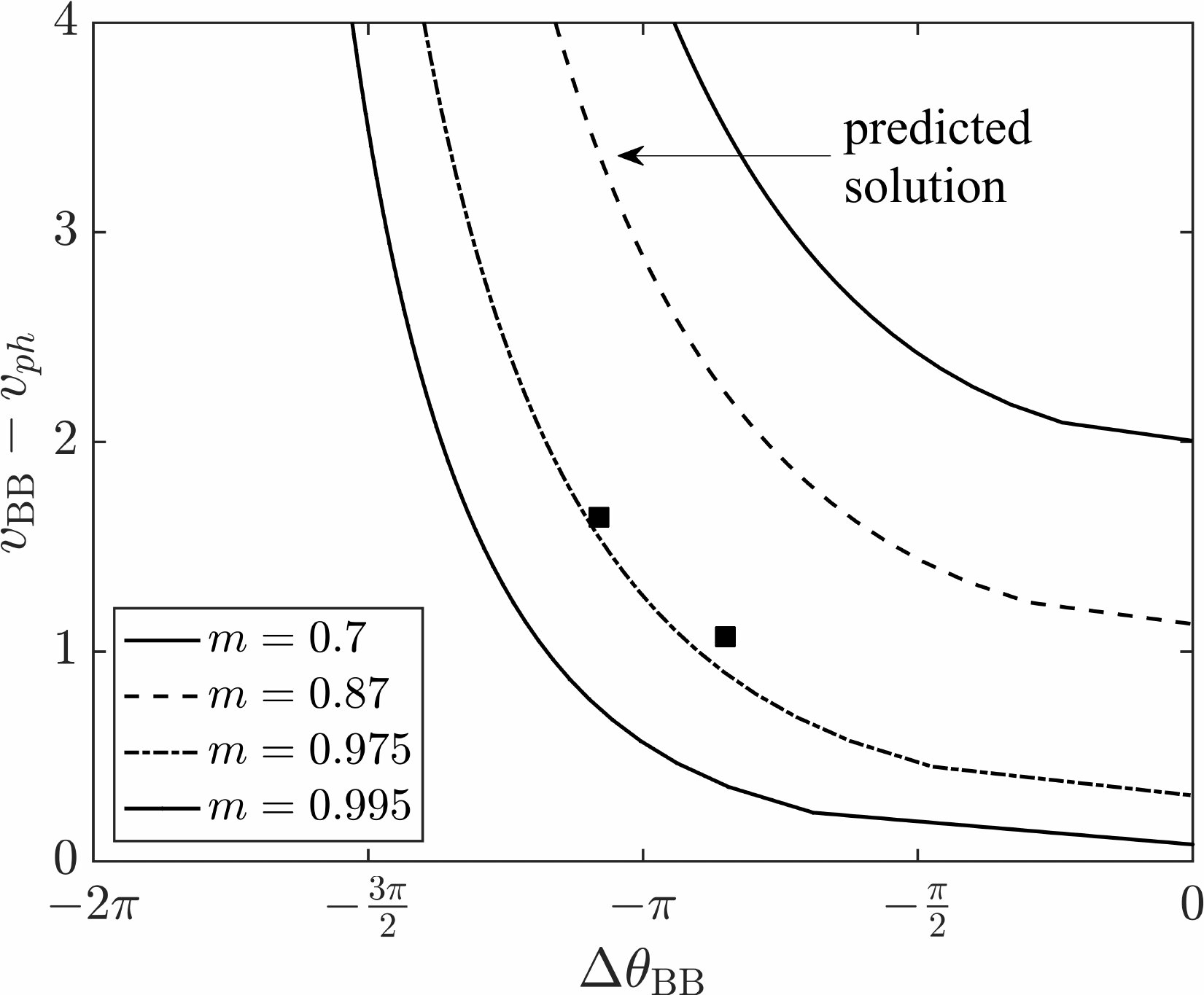}}
    \hspace{0.1in}
    \subfloat[]{\includegraphics[height=0.35\textwidth]{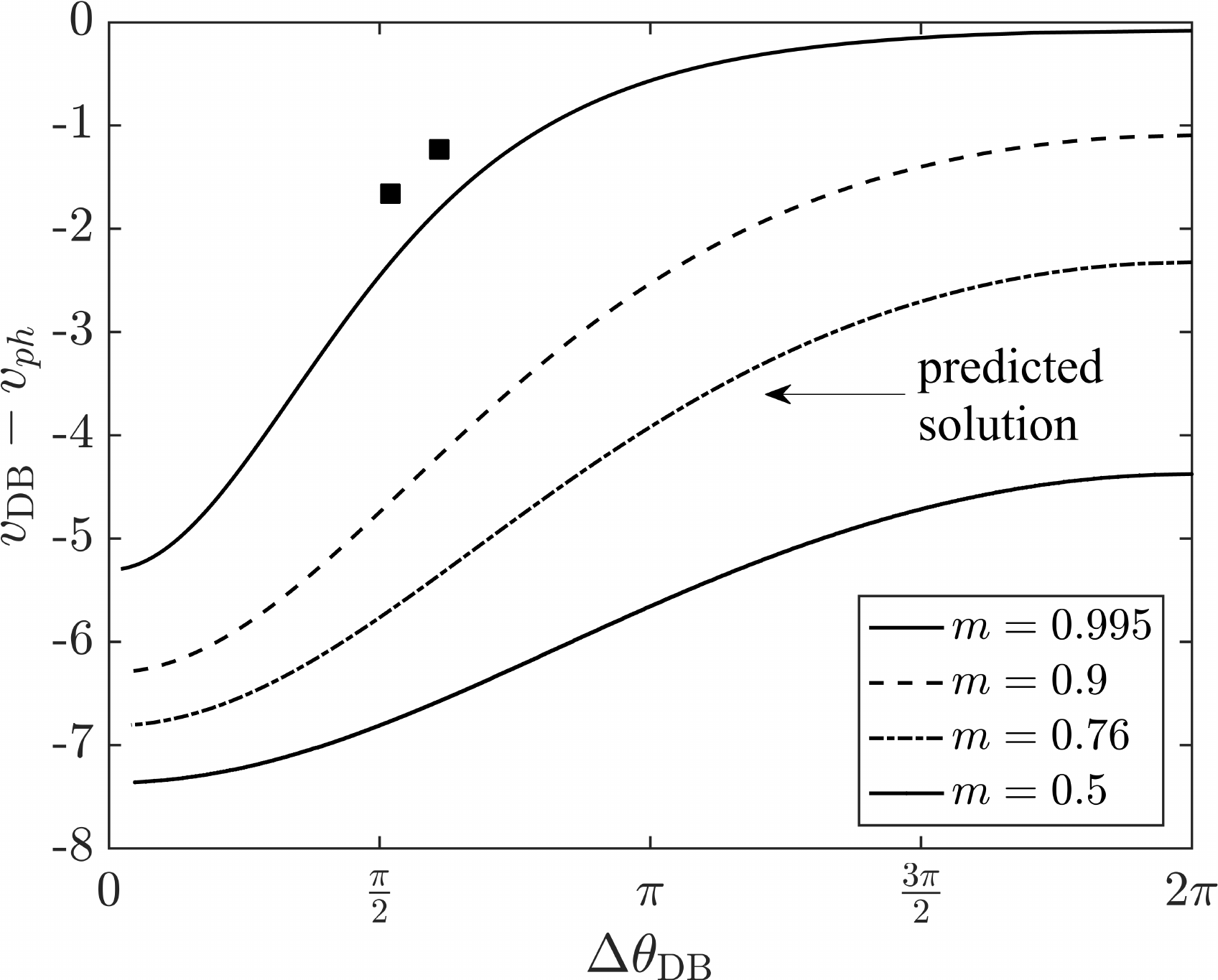}}
    \caption{Representative KdV traveling breather nonlinear dispersion relation curves for the BB (left) and DB (right) solutions. Markers are  the measured conduit BB (left) and DB (right) speed-phase shift relation.}
    \label{fig:KdV_disp}
\end{figure}

The nonlinear dispersion relation of the KdV traveling breathers solutions is given by \cite{hoefer2022KDV}
\begin{align*}
    \alpha_b = F(\varphi_\gamma,m), \quad \kappa_b=\frac{\sqrt{1-\lambda-m}\sqrt{-\lambda-m}}{\sqrt{1-2m-\lambda}} - Z(\varphi_\gamma,m), \quad c_b = c_0 + \frac{4\sqrt{1-\lambda-2m}\sqrt{1-\lambda-m}\sqrt{-\lambda-m}}{\kappa_b}
\end{align*}
for the BB, where $\varphi_\gamma\in(0,\frac{\pi}{2})$ is found from $\sin \varphi_\gamma = \frac{\sqrt{-\lambda-m}}{\sqrt{1-2m-\lambda}}$, and 
\begin{align*}
    \alpha_d = F(\varphi_\alpha,m), \quad \kappa_d=Z(\varphi_\alpha,m), \quad c_d = c_0 - \frac{4\sqrt{\lambda+m}\sqrt{\lambda-1+2m}\sqrt{1-\lambda-m}}{\kappa_d}
\end{align*}
for the DB, where $\varphi\in(0,\frac{\pi}{2})$ is from $\sin \varphi_\alpha=\frac{\sqrt{1-m-\lambda}}{\sqrt{m}}$. The normalized phase shift of the background cnoidal wave can be written as 
\begin{equation*}
    \Delta \theta_{\rm BB} = -\frac{2\pi \alpha_b}{K(m)} \in (-2\pi,0], \quad 
    \Delta \theta_{\rm DB} = \frac{2\pi \alpha_d}{K(m)} \in [0,2\pi).
\end{equation*} 
Figure \ref{fig:KdV_disp} demonstrates the dependence of breather-phase speed difference versus breather normalized phase shift of the KdV breather solutions.
We then compare the BB-BB and DB-DB interaction experiments in Fig.4 of the main text with the KdV traveling breather nonlinear dispersion relation in Fig.\ref{fig:KdV_disp}. The BB-BB interaction in Fig.4a of the main text is on a nondimensional carrier background of $(a,\bar{A},k)=(0.96,1.00,0.39)$ so that $m=0.87$. The measured cnoidal phase speed is $v_{ph}=1.74$, while the predicted is $C_0=1.81$. The BB speeds and phase shifts are $(v_1,\Delta\theta_1)=(3.38,-1.08\pi)$, $(v_2,\Delta\theta_2)=(2.81,-0.85\pi)$. In Fig.~\ref{fig:KdV_disp}a, we plot the observed BBs' breather-phase speed differences against their phase shifts. The predicted solution corresponds to the curve at $m=0.87$, but we observe that the experimental BB lies on the curve at $m=0.975$. This means that if we fix the carrier background for the same phase shift, the experimental BB has a smaller breather-phase speed difference than the approximate KdV BB. Nevertheless, our conduit BB-BB interaction experiment agrees qualitatively with the KdV traveling breather nonlinear dispersion relation since an increasing BB speed corresponds to an increasing BB amplitude
and \textit{decreasing} BB phase shift.

On the other hand, the DB-DB interaction in Fig.4c of the main text is on a carrier of $(a,\bar{A},k)=(2.98,0.36,2.31)$, implying $m=0.76$. Again, we obtain the measured and predicted carrier phase speeds $v_{ph}=3.39$ and $C_0=4.71$, respectively. The measured DB properties are  $(v_1,\Delta\theta_1)=(2.16,0.61\pi)$, $(v_2,\Delta\theta_2)=(1.73,0.52\pi)$. In Fig.~\ref{fig:KdV_disp}b, we find the speed-phase shift relation of the experimental DBs is close to the KdV nonlinear dispersion relation at $m=0.995$. Qualitative agreement is achieved with the KdV prediction since an increasing DB speed corresponds to an increasing DB phase shift.

